\documentclass[
 reprint,
 showpacs, preprintnumbers,
 amsmath, amssymb,
 aps,
 prl,
 floatfix,
 superscriptaddress
]{revtex4-2}

\usepackage{amsmath}
\usepackage{graphicx}
\usepackage{dcolumn}
\usepackage{bm}
\usepackage{bbm}

\usepackage[utf8]{inputenc}
\usepackage[T1]{fontenc}
\usepackage{mathptmx}
\usepackage{textgreek}
\usepackage{upgreek}
\usepackage{float}
\usepackage{hyperref}

\usepackage{color}

\usepackage{natbib}

\newcommand{\beginsupplement}{%
        \setcounter{table}{0}
        \renewcommand{\thetable}{S\arabic{table}}%
        \setcounter{figure}{0}
        \renewcommand{\thefigure}{S\arabic{figure}}%
				\renewcommand{\theequation}{S.\arabic{equation}}
     }
		
\begin{document}

	\preprint{}

\title[Diode reversal]{Sign reversal of the AC and DC supercurrent diode effect and  0--$\pi$-like transitions in ballistic Josephson~junctions} 



\author{A.~Costa$^\dagger$}
\affiliation{Institut f\"ur Theoretische Physik, University of Regensburg, 93040 Regensburg, Germany}
\author{C.~Baumgartner$^\dagger$}
\author{S.~Reinhardt}
\author{J.~Berger}
\affiliation{Institut f\"ur Experimentelle und Angewandte Physik, University of Regensburg, 93040 Regensburg, Germany}

\author{S.~Gronin}
\author{G.~C.~Gardner}
\affiliation{Birck Nanotechnology Center, Purdue University, West Lafayette, Indiana 47907 USA}
\author{T.~Lindemann}
\affiliation{Birck Nanotechnology Center, Purdue University, West Lafayette, Indiana 47907 USA}
\affiliation{Department of Physics and Astronomy, Purdue University, West Lafayette, Indiana 47907 USA}

\author{M.~J.~Manfra}
\affiliation{Birck Nanotechnology Center, Purdue University, West Lafayette, Indiana 47907 USA}
\affiliation{Department of Physics and Astronomy, Purdue University, West Lafayette, Indiana 47907 USA}
\affiliation{School of Materials Engineering, Purdue University, West Lafayette, Indiana 47907 USA}
\affiliation{Elmore Family School of Electrical and Computer Engineering, Purdue University, West Lafayette, Indiana 47907 USA}

\author{D.~Kochan}
\affiliation{Institut f\"ur Theoretische Physik, University of Regensburg, 93040 Regensburg, Germany}
\affiliation{Institute of Physics, Slovak Academy of Sciences, 84511 Bratislava, Slovakia}

\author{J.~Fabian}
\affiliation{Institut f\"ur Theoretische Physik, University of Regensburg, 93040 Regensburg, Germany}

\author{N.~Paradiso}\email{nicola.paradiso@physik.uni-regensburg.de}
\author{C.~Strunk}
\affiliation{Institut f\"ur Experimentelle und Angewandte Physik, University of Regensburg, 93040 Regensburg, Germany}
%


\begin{abstract}	
The recent discovery of intrinsic supercurrent diode effect, and its prompt observation in a rich variety of systems, has shown that nonreciprocal supercurrents naturally emerge when both space- and time-inversion symmetries are broken. In Josephson junctions, nonreciprocal supercurrent can be conveniently described in terms of spin-split Andreev states. Here, we demonstrate a sign reversal of the supercurrent diode effect, in both its AC and DC manifestations. In particular, the AC diode effect –-i.e., the asymmetry of the Josephson inductance as a function of the supercurrent—- allows us to probe the current-phase relation near equilibrium. Using a minimal theoretical model, we can then link the sign reversal of the AC diode effect to the so-called $0-\pi$-like transition, a predicted, but still elusive feature of multi-channel junctions. Our results demonstrate the potential of inductance measurements as sensitive probes of the fundamental properties of unconventional Josephson junctions.

\end{abstract}

\maketitle

\section{Introduction}

Recent experiments~\cite{Ando2020,Baumgartner2022,BaumgartnerSI2022,Wu2022,Jeon2022,Pal2022,Bauriedl2022,Turini2022,FrolovQW2022} have shown that it is possible to obtain nonreciprocal supercurrents by solely acting on the spin degree of freedom of a superconductor through Zeeman fields. Since then, supercurrent rectifiers---called superconducting diodes---have generated keen interest owing to their potential applications in dissipation-free electronics. From a fundamental point of view, the superconducting diode effect proved to be an important probe of symmetry breaking in novel and exotic superconducting systems, such as magic-angle twisted bi-~\cite{diezmerida2021magnetic} or trilayer~\cite{lin2022zerofield,scammell2022theory} graphene, as well as proximitized topological insulators~\cite{BoLu2022,Pal2022,Fu2022TopoDiode}.

Spin-orbit interaction~(SOI) is the key ingredient to obtain supercurrent nonreciprocity through a Zeeman 
coupling, as first demonstrated  for both  bulk superconductors
and Josephson junctions~\cite{Baumgartner2022,BaumgartnerSI2022}. The phenomenology in both cases is similar: supercurrent rectification is induced by a Zeeman splitting of electron states with opposite momentum, which can happen in systems with spin-momentum locking due to the underlying SOI.


To date, theoretical models use different approaches to describe the supercurrent diode effect (SDE) in bulk superconductors and Josephson junctions. 
In the former case, nonreciprocity is usually attributed to the emergence of a helical phase~\cite{daido2021prl,yuan2021pnas,he2021njp,Ilic2022}, i.e., to the finite Cooper-pair momentum that results from the
SOI-split Fermi surface shifted due to Zeeman coupling. 
Instead, nonreciprocal supercurrent in superconducting-normal-superconducting~(S-N-S) Josephson junctions is more conveniently described in terms of Andreev bound states~\cite{Andreev1966,*Andreev1966alt} that, in the presence of Zeeman coupling and SOI, modify the current-phase relation (CPR) endowing it with
an \textit{anomalous} phase shift~$ \varphi_0 $, with~$ \varphi_0 \neq 0 , \pi $~\cite{Bezuglyi2002,Krive2004,Buzdin2008,Reynoso2008,Zazunov2009,Liu2010a,Liu2010,Liu2011,Reynoso2012,YokoyamaJPSJ2013,Brunetti2013,Yokoyama2014,Shen2014,Yokoyama2014,Konschelle2015,Szombati2016,Assouline2019,Mayer2020b,Strambini2020}. Such phase shift leads to a marked asymmetry of the current-phase relation (CPR), i.e., $ I(-\varphi) \neq -I(\varphi) $.
In the simplest case, the 
magnitude of $ \varphi_0 $ depends on the products of SOI and Zeeman coupling strengths, and the inverse of velocity square of the impinging electrons~(holes)~\cite{Buzdin2008}.
Since the latter is different for each transverse channel in the junction, the CPRs of individual channels acquire different $ \varphi_0 $-shifts. The last ingredient necessary to obtain the diode effect is a skewed CPR, i.e., a CPR with higher harmonics, as those observed in short-ballistic junctions. This requirement arises from the fact that the sum of different (also shifted) sine-functions is still sinusoidal, hence positive and negative critical current have the same absolute value.
Instead, the sum of skewed CPRs with different $ \varphi_0 $-shifts produces a \textit{distortion} of the total CPR that breaks the symmetry between its positive and negative branch.
We shall refer to the resulting difference between positive and negative critical current as the DC supercurrent diode effect (SDE).
The asymmetry between positive and negative branch of the CPR
also implies that its inflection point shifts to finite current $i^{\ast}$. Such inflection point current corresponds to the minimum of the Josephson inductance $L$ measured as a function of the current $I$. 
The finite $i^{\ast}$ renders $L(I)$ asymmetric around $I=0$, reflecting the magnetochiral anisotropy in the Josephson inductance~\cite{Baumgartner2022}. Thus, the impedance for small AC signals will depend on the polarity of $I$, an effect that can be seen as the AC counterpart of the DC SDE.  In what follows, we shall refer to the asymmetry in critical current and to that in $L(I)$, respectively, as the DC and AC SDE.

\begin{figure*}[tb]
\centering
\includegraphics[width=\textwidth]{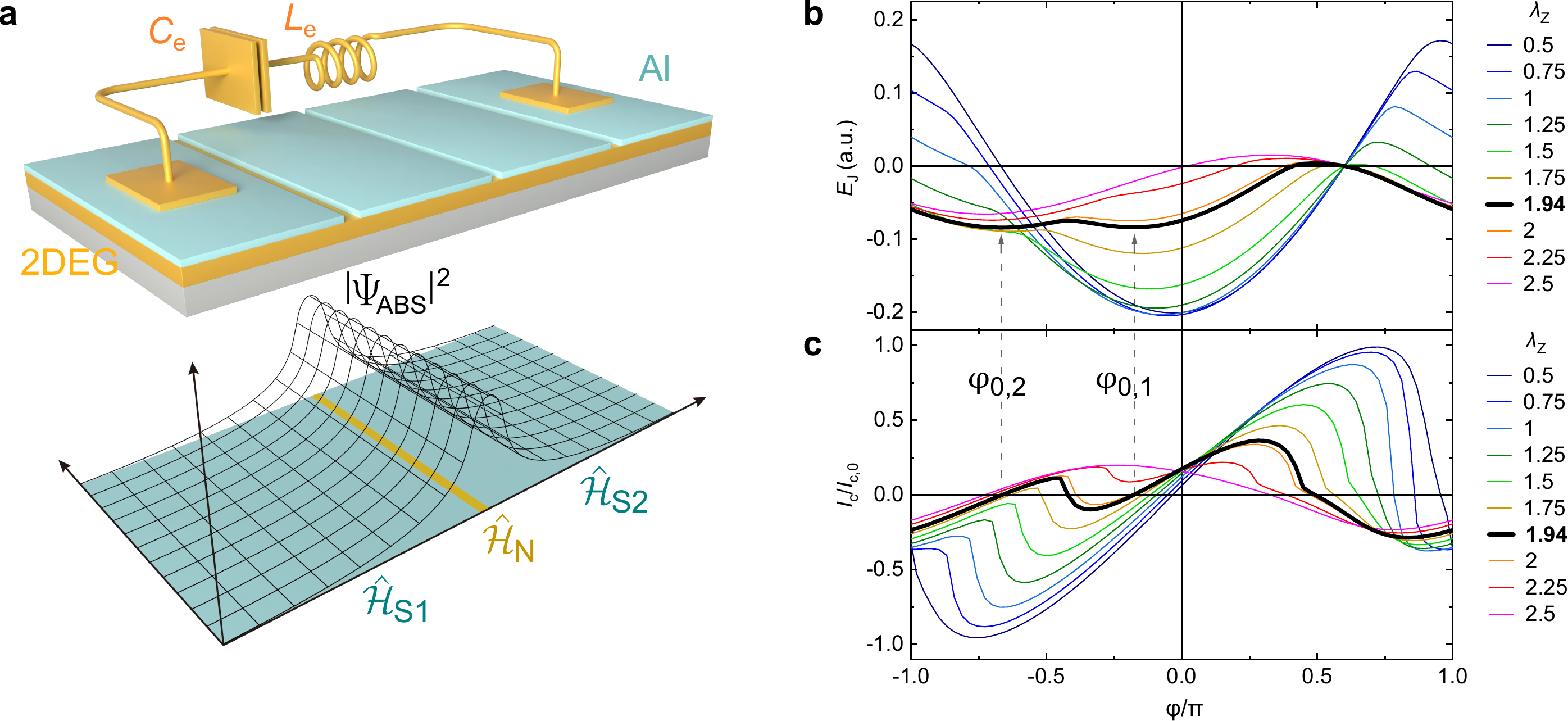}
\caption{\textbf{Multi-channel Rashba Josephson junctions: experiment and theory.} \textbf{a}, (Top)~Sketch of the device, consisting of a 1D Josephson junction array in series with a LC circuit. The superconducting regions consist of a 2D electron gas~(2DEG; gold) proximitized by epitaxially grown Al islands~(turquoise); they are connected through 2DEG weak links that are indicated by the gaps between the Al islands, see text. (Bottom)~Theoretical model: two 2D~Rashba superconductors~(turquoise), described by the Hamiltonians~$ \hat{\mathcal{H}}_{\mathrm{S},1} $ and $ \hat{\mathcal{H}}_{\mathrm{S},2} $, are connected by a deltalike N-link~(gold) described by the Hamiltonian~$ \hat{\mathcal{H}}_\mathrm{N} $; see Methods. The Andreev bound state wave~function~$ \Psi_\mathrm{ABS} $ (black lines show its absolute square) is strongly localized around the barrier.
\textbf{b}, Computed Josephson~energy as a function of the phase difference~$ \varphi $ for various indicated Zeeman~parameters~$ \lambda_\mathrm{Z} $; see text. Arrows indicate the positions of the two relevant energy~minima, $ \varphi_\mathrm{0,2} $ and $ \varphi_\mathrm{0,1} $, which become degenerate at~$ \lambda_\mathrm{Z}=1.94 $~(thick black curve) when the 0--$\pi$-like transition occurs. 
\textbf{c}, Computed current-phase relations for the Zeeman~parameters considered in~\textbf{b}; the current is normalized to the critical current 
at zero magnetic field, $I_\mathrm{c,0} \equiv I_\mathrm{c}( \lambda_\mathrm{Z}=0)$.}
\label{fig:first}
\end{figure*}

In this work, we demonstrate that large Rashba SOI together with a Zeeman field give rise to a reversal of the AC SDE in ballistic multi-channel Josephson junctions. 
We theoretically show that all experimental observations can be unambiguously explained by a minimal model linking the supercurrent flow to the underlying short-junction Andreev bound states. We demonstrate that the reversal of the AC SDE naturally emerges as a consequence of the so-called $ 0 $--$ \pi $-like transition, where the Josephson junction switches between two distinct minima of the corresponding energy-phase relation. Such transitions were predicted nearly a decade ago~\cite{Yokoyama2014}, but, to the best of our knowledge, not yet observed in experiments.

In our experiments, we also observe a sign reversal of the DC SDE, i.e., of the critical current difference for the two bias polarities as a function of the in-plane field. This reversal occurs at an in-plane field close to the one delimiting the  $ 0 $--$ \pi $-like transition. With the help of our theoretical model, we can show that the reversal of both the AC and DC SDE originate from the peculiar shape of the CPR in Rashba Josephson junctions under large in-plane fields.


\section{Device and model description}





We briefly describe our experiment, which is discussed in detail in the Methods~section, as well as, e.g., in Refs.~\cite{baumgartner2020,Baumgartner2022}. 
The starting point is a heterostructure featuring a very shallow InAs/InGaAs quantum well, hosting a 2D electron gas (2DEG) with large Rashba SOI. The heterostructure is capped by a 7-nm-thick epitaxial layer of Al, which induces superconductivity in the 2DEG by proximity effect. By deep-etching, we define a mesa with a width of 3.15~\textmu m for sample~1 and 3.27~\textmu m for sample~3, respectively. By electron-beam lithography, we then selectively etch the Al film so that a gap of 100~nm is left between the adjacent pristine Al islands. We define an array of 2250 islands, with periodicity 1.1 \textmu m~\cite{baumgartner2020,Baumgartner2022}. In the top part of Fig.~\ref{fig:first}\textbf{a}, we sketch our device, showing only a few junctions for clarity. The reason why we work with long arrays and not with single junctions is merely technical, and it is discussed in the Supplementary Information. The 2DEG in the gap regions acts as a ballistic normal weak link connecting the superconducting 2DEG portions,  in which the Al film induces a sizable gap $\Delta^{\ast}=130$~\textmu eV, i.e., approximately 60\% of that of the epitaxial Al~\cite{baumgartner2020}. The transparency of our junctions is typically very high~\cite{baumgartner2020}, with average transmission coefficients of $\bar{\tau}=0.94$ for sample~1 and $\bar{\tau}=0.93$ for sample~3, respectively.

In our theoretical model, a short 2D S-N-S~junction consists of two semi-infinite Rashba $ s $-wave superconductors~(S) phase-coherently coupled by a thin nonsuperconducting~(N) link~(see~Fig.~\ref{fig:first}\textbf{a}, bottom).
For simplicity, the short junction is described by a deltalike tunneling barrier, which also contains the Zeeman coupling due to the applied in-plane magnetic field. The Zeeman coupling is parameterized by $ \lambda_\mathrm{Z} = 2 m V_\mathrm{Z} d / (\hbar^2 k_\mathrm{F}) $, 
with the Zeeman potential~$V_\mathrm{Z} $, the effective electron mass~$ m $, the weak link thickness~$ d $, and the 2DEG Fermi wave vector~$ k_\mathrm{F} $; see Methods.
Magnetic field  effects inside the bulk of the S~regions~(in particular the suppression of the proximity-induced superconducting gap with increasing magnetic field) are neglected.


Coherent Cooper-pair tunneling through the S-N-S junction is mediated by Andreev bound states that are localized around the N~link~(see~Fig.~\ref{fig:first}\textbf{a}, bottom). To determine their energies and wave functions, we solve the stationary Bogoliubov--de~Gennes~equation~\cite{DeGennes1989}. From the Andreev spectrum 
we obtain the phase-asymmetric CPR,~$ I(\varphi) $, and consequently the polarity-dependent critical currents~$ I_\mathrm{c}^+ $ and $ I_\mathrm{c}^- $, as well as the Josephson inductance 
    \begin{equation}
        L(I) = \frac{\hbar}{2e} \frac{\mathrm{d} \varphi(I)}{\mathrm{d} I} ,
    \end{equation}
to be compared with the experimental data.

Figure~\ref{fig:first}\textbf{b} illustrates the Josephson energy-phase relations, $E_{\text{J}}(\varphi)$, evaluated from our theoretical model for different Zeeman~parameters~$ \lambda_\mathrm{Z} \in [0.5;2.5] $. Note that the energy-phase relation in a multi-channel system is given by
$E_\mathrm{J}(\varphi)=\sum_i\varepsilon_i(\varphi)$, where the sum runs over the individual transverse channels.
Our inductance measurements always probe the system in the vicinity of the global minimum of $E_\mathrm{J}(\varphi)$, which is the experimental working point. Since $\varphi$ at which $E_\mathrm{J}(\varphi)$ is minimal is simultaneously a zero of $I(\varphi)$, such phase, by definition, corresponds to the anomalous phase shift $\varphi_0$.

The main results of our analytical calculations, see Fig.~\ref{fig:first}\textbf{b}, are that (i)~$E_\mathrm{J}(\varphi)$ features, at sufficiently high Zeeman fields $\lambda_\mathrm{Z}$, two minima---one global and one local; (ii) when increasing $\lambda_\mathrm{Z}$, the lower (global) minimum increases in energy, while the upper (local) minimum decreases, until  a degeneracy point is reached where the system switches from one minimum (i.e., from one anomalous phase $\varphi_{0,1}$) to the other ($\varphi_{0,2}$). In Fig.~\ref{fig:first}\textbf{b}, this transition occurs for $\lambda_\mathrm{Z} = 1.94$ (black curve). 
Unlike for conventional $0$--$\pi$  transitions~\cite{LiPRL2019,Hart2017,ChenNatComm2018,Ke2019,WhiticarPRB2021}, the phase difference $\Delta \varphi \equiv |\varphi_\mathrm{0,1}- \varphi_\mathrm{0,2}|$ is significantly less than $\pi$, owing to the anomalous phase shift $\varphi_0$. As discussed below, one experimental signature of this transition is the reversal of the AC SDE.

\section{Results and interpretation}

\begin{figure*}[htb]
\centering
\includegraphics[width=\textwidth]{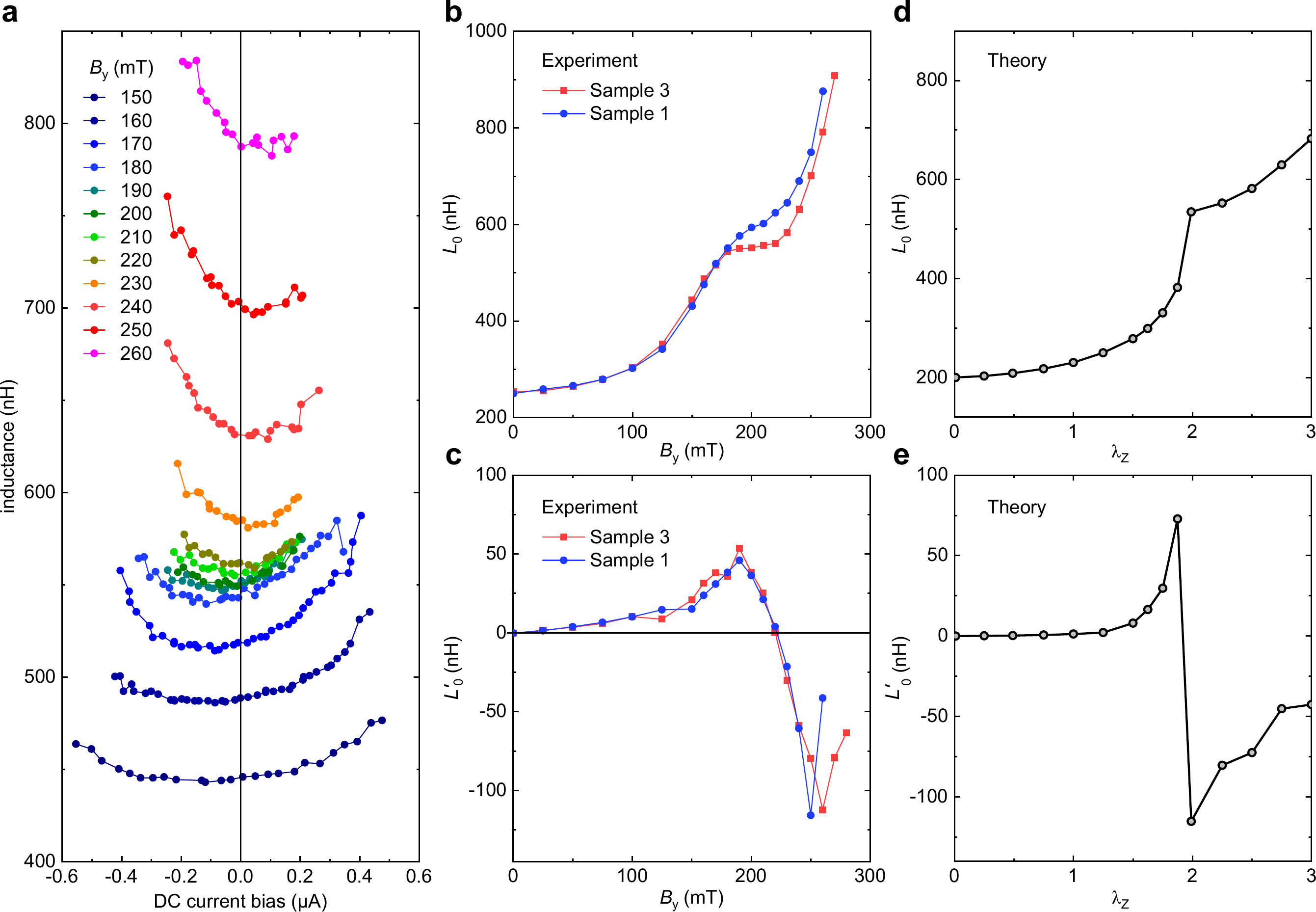}
\caption{
\textbf{Reversal of the AC supercurrent diode effect at the $ 0 $--$\pi$-like transition.} 
\textbf{a}, Josephson inductance~$ L(I) $ for the array of sample~3, measured at $T=100$~mK as a function of the DC current~$I$ and for different values of the in-plane field component $B_y$ perpendicular to the current. \textbf{b}, Constant term~$L_0$ of the polynomial expansion of $L(I)$, plotted versus $B_y$. \textbf{c}, Linear coefficient~$L^{\prime}_0$ of the polynomial expansion of $L(I)$, plotted versus $B_y$. Red (blue) symbols refer to sample~3 (sample~1). 
\textbf{d}, $L_0$ as computed from our theoretical model as a function of the phenomenological Zeeman~parameter~$ \lambda_\mathrm{Z} $. \textbf{e}, $L^{\prime}_0$ as computed from our theoretical model as a function of $ \lambda_\mathrm{Z} $. }  
\label{fig:second}
\end{figure*}

Figure~\ref{fig:second}\textbf{a} shows our main experimental results, namely, the in-plane field-induced reversal of the AC SDE.  In the graph, we report the Josephson inductance $L$ as a function of the DC current $I$ for different values of the in-plane magnetic field $B_y$, measured on sample~3. At moderate fields ($B_y<200$~mT), the $L(I)$-curves are asymmetric around zero bias, with a minimum occurring at current $i^{\ast}<0$, which corresponds to the inflection point of the CPR~\cite{baumgartner2020,Baumgartner2022}. 

Our crucial observation is the fact that at higher fields ($B_y>200$~mT), the sign of $i^{\ast}$ is inverted.  
By Taylor expansion of the $L(I)$-curves to the first order in $I$, we get two coefficients, i.e.,~$L_0\equiv L(0)$ and $L^{\prime}_0\equiv \partial_I L(0)$, that as functions of the magnetic field serve as figures of merit of the AC SDE~\cite{Baumgartner2022}. 
In Figs.~\ref{fig:second}\textbf{b} and \textbf{c}, we show the $B_y$-field dependence of $L_0$ and $L^{\prime}_0$ for sample~3~(red symbols, corresponding to the data in Fig.~\ref{fig:second}\textbf{a}) and for sample~1~(blue symbols), respectively. We notice that (i)~the $L_0(B_y)$-curves feature a plateau at about 180~mT, corresponding to the accumulation of the $L(I)$-curves in Fig.~\ref{fig:second}\textbf{a}; (ii)~$L^{\prime}_0$ shows a nearly linear increase at low $B_y $-fields followed by an upturn for $B_y >100$~mT, with a peak at about 180~mT; finally (iii)~a dramatic drop with sign change occurs  at $B_y=220$~mT, reflecting the sign change of $i^{\ast}$ in panel~\textbf{a}, which is our main experimental finding.

\begin{figure*}[htb]
\centering
\includegraphics[width=\textwidth]{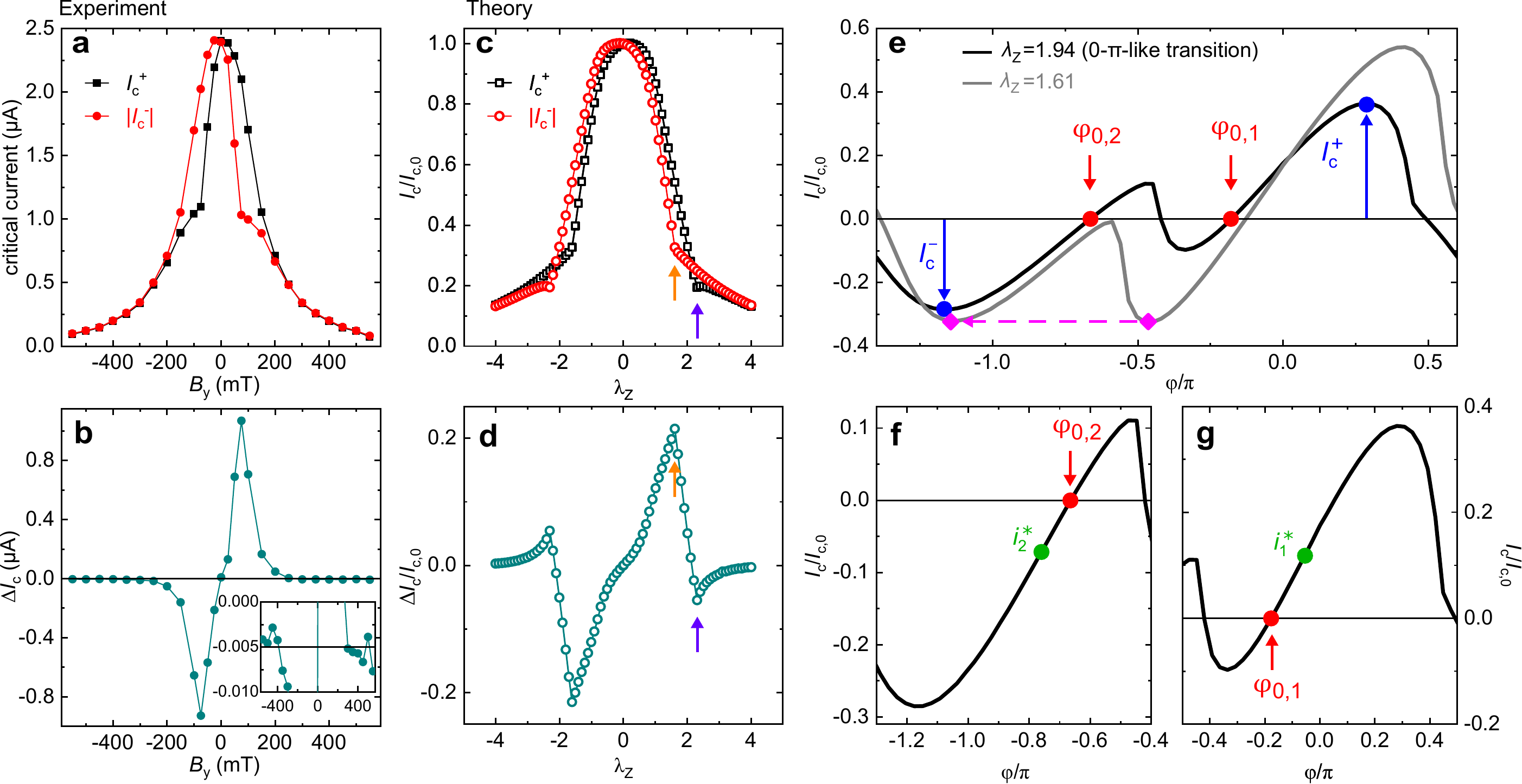}
\caption{
\textbf{Reversal of the DC supercurrent diode effect.} 
\textbf{a}, Absolute value of the critical current measured on sample 1 for positive ($I_\mathrm{c}^+$) and negative ($|I_\mathrm{c}^-|$) bias, plotted as a function of the in-plane field $B_y$ perpendicular to the current. \textbf{b}, Critical-current difference~$\Delta I_\mathrm{c}=I_\mathrm{c}^+-|I_\mathrm{c}^-|$. The inset shows a zoom-in on a reduced range in the vertical axis. 
\textbf{c}, $I_\mathrm{c}^+$ and $|I_\mathrm{c}^-|$ as computed from our theoretical model as a function of the phenomenological Zeeman~parameter~$ \lambda_\mathrm{Z} $. The critical current values are normalized to that for~$ \lambda_\mathrm{Z} = 0 $, i.e., $I_{\mathrm{c},0}\equiv I_\mathrm{c}(\lambda_\mathrm{Z} = 0)$. For $ \lambda_\mathrm{Z} = 1.61$ (orange arrow) and $ \lambda_\mathrm{Z} = 2.31$ (purple arrow), two CPR minima and two CPR maxima become degenerate. This corresponds to a discontinuity in the slope of $|I_\mathrm{c}^-(B_y)|$ and $I_\mathrm{c}^+(B_y)$, respectively (see text). 
\textbf{d}, Critical current difference~$\Delta I_\mathrm{c} (\lambda_Z)$ normalized to $I_\mathrm{c,0}$, as deduced from the previous panel. 
\textbf{e}, Computed current-phase relation~(CPR) for $ \lambda_\mathrm{Z}=1.94$~(black), corresponding to the $0$--$\pi$-like transition in our model, and for $ \lambda_\mathrm{Z}=1.61$~(gray), corresponding to the equality of the two minima of the CPR. For the $ \lambda_\mathrm{Z}=1.94$-curve, red dots correspond to the anomalous phase shifts, while blue dots correspond to the absolute maximum and minimum of the CPR. For the $ \lambda_\mathrm{Z}=1.61$-curve, the magenta arrow indicates the transition from one CPR minimum to another (magenta diamonds).
\textbf{f}, Zoom-in near the second anomalous phase $\varphi_{0,2}$ for the $ \lambda_\mathrm{Z}=1.94$-curve in panel~\textbf{e}. The green dot indicates the first inflection point $i^{\ast}_2<0$, which is a local minimum of the Josephson inductance~$L(I)$.
\textbf{g}, Zoom-in near the first anomalous phase $\varphi_{0,1}$ of the $ \lambda_\mathrm{Z}=1.94$-curve in panel~\textbf{e}. Note the opposite sign of $i^{\ast}_1$ compared to $i^{\ast}_2$.
}
\label{fig:third}
\end{figure*}

Figures~\ref{fig:second}\textbf{d} and \textbf{e} show the  corresponding results of our theoretical calculations for $ L_0(\lambda_\mathrm{Z}) $ and $L^{\prime}_0(\lambda_\mathrm{Z}) $, respectively, as functions of $\lambda_\mathrm{Z}$. 
Although  our analytical model contains only a single parameter ($\lambda_\mathrm{Z}$), it captures even quantitatively all the relevant experimental observations~(i)--(iii). In particular, the plateau for~$ L_0 $ and the $ L^{\prime}_0 $-peak followed by a sudden sign~change are visible near $\lambda_\mathrm{Z} = 2$ in Fig.~\ref{fig:second}\textbf{d} and Fig.~\ref{fig:second}\textbf{e}, respectively. In our calculations, both effects 
reflect the $ 0 $--$ \pi $-like~transition, where the anomalous phase switches from~$ \varphi_\mathrm{0,1}$ to~$ \varphi_\mathrm{0,2} $, see Fig.~\ref{fig:first}\textbf{c}. In the vicinity of $ \varphi_\mathrm{0,1}$  and $ \varphi_\mathrm{0,2} $,  the sign of the CPR curvature (i.e.,  $\mathrm{d}^2 I/\mathrm{d}\varphi^2$) differs, see Fig.~\ref{fig:third}\textbf{e-g}. Correspondingly, the  sign   of $L^{\prime}_0\sim \mathrm{d}^2\varphi/\mathrm{d}I^2$ in the measurements changes as well.


Complementary to the AC SDE is the DC SDE, i.e., the polarity-dependent critical current.
Figure~\ref{fig:third}\textbf{a} shows the $B_y$-dependence of the critical current measured on sample~1 for the two bias polarities, namely, $I_\mathrm{c}^+$ and $|I_\mathrm{c}^-|$. 
We observe an evident critical current asymmetry, i.e.~a pronounced DC SDE. This can be better seen in Fig.~\ref{fig:third}\textbf{b}, where we plot the difference $\Delta I_\mathrm{c} \equiv I_\mathrm{c}^+-|I_\mathrm{c}^-|$. The main features of this graph are the nearly-linear $B_y$-dependence of $\Delta I_\mathrm{c}$ up to 75~mT followed by a sharp suppression of the rectification at higher values. A secondary, yet evident, feature is a slight shoulder 
at zero field. These characteristics of $\Delta I_\mathrm{c}$ were already reported in Ref.~\cite{Baumgartner2022}. In the inset of Fig.~\ref{fig:third}\textbf{b}, we show a zoom-in of the $\Delta I_\mathrm{c}(B_y)$-curve. We observe that $\Delta I_\mathrm{c}$ converges to a finite value of $-5$~nA at high fields---for both $B_y>0$ and $B_y<0$---which we attribute to an instrumental offset. After subtracting this offset, we see that the graph of $\Delta I_\mathrm{c}$ is point-symmetric around the origin  and it changes sign at $|B_y|\approx 400$~mT, before converging to zero. 
It appears natural to ask how such a reversal of the 
DC SDE is related to the AC SDE displayed in Fig.~\ref{fig:second}. We notice that the reversal of $\Delta I_\mathrm{c}$ is very weak, at the limit of visibility in our experiment, as opposed to the strong inversion of $L^{\prime}_0$ in Fig.~\ref{fig:second}\textbf{c}.

The results of our analytical calculations help to clarify the origin of the slope discontinuity at $B_y=75$~mT (-75~mT) for $|I_\mathrm{c}^-|$ ($I_\mathrm{c}^+$), see Fig.~\ref{fig:third}\textbf{a}. This slope discontinuity is responsible for the sharp $\Delta I_\mathrm{c}$ suppression in Fig.~\ref{fig:third}\textbf{b}, and eventually for its sign reversal. Figure~\ref{fig:third}\textbf{c} presents the computed critical current for positive~($I_\mathrm{c}^+$) and negative~($|I_\mathrm{c}^-|$) bias as a function of the Zeeman parameter~$\lambda_\mathrm{Z}$, while Fig.~\ref{fig:third}\textbf{d} shows their respective difference~$\Delta I_\mathrm{c}$. In both cases, the values are normalized to the critical current at $\lambda_\mathrm{Z}=0$. 
Despite its simplicity, the minimal model describes the main features of the experimental DC SDE, including the shoulder near $B_y=0$. The computed  $\Delta I_\mathrm{c}$ reaches a maximum 
at $ \lambda_\mathrm{Z} \approx 1.61 $~(marked by the orange arrows in Figs.~\ref{fig:third}\textbf{c}~and~\textbf{d})
and experiences a discontinuous change in its slope that is followed by a steep descent to negative values. The slope of $\Delta I_\mathrm{c}$ changes again at $ \lambda_\mathrm{Z} \approx 2.31 $~(marked by the purple arrows in Figs.~\ref{fig:third}\textbf{c}~and~\textbf{d}), where the supercurrent rectification reaches its opposite maximum, 
before it eventually converges to zero at higher fields. 

For an intuitive picture of the physical mechanism leading to the reversal of the DC and AC SDEs, we need to look at the CPR in a large Zeeman field.
In Fig.~\ref{fig:third}\textbf{e}, black curve, we display the computed CPR for $\lambda_Z=1.94$, which corresponds to the $ 0 $--$ \pi $-like transition. As discussed above~(see Figs.~\ref{fig:first}\textbf{b,c}), the unbiased system switches  from the working point~$\varphi_\mathrm{0,1}$ to $\varphi_\mathrm{0,2}$ at that value. For low current bias~(as it is the case for AC SDE experiments), the CPR probed near $\varphi_\mathrm{0,1}$ is very different from that measured near $\varphi_\mathrm{0,2}$, as shown in the zoom-in graphs displayed in Fig.~\ref{fig:third}\textbf{g} and  Fig.~\ref{fig:third}\textbf{f}, respectively. 
Both portions of the CPR are strongly asymmetric, but the magnitude and sign of the asymmetry are different. For instance, to reach the minimum of the $L(I)$-curve (namely, the inflection point $i^{\ast}$ of the CPR), one needs to apply a positive bias near $\varphi_\mathrm{0,1}$~(Fig.~\ref{fig:third}\textbf{g}, green dot), whereas the bias must be negative to reach the inflection point near $\varphi_\mathrm{0,2}$, i.e.,  after the $ 0 $--$ \pi $-like transition~(Fig.~\ref{fig:third}\textbf{f}). In other words, the second derivatives (i.e.,~the curvature) of the CPR  in $\varphi_\mathrm{0,1}$ and $\varphi_\mathrm{0,2}$ have opposite signs. As a consequence, the computed $L_0^{\prime}\sim \mathrm{d}^2\varphi/dI^2$ must have a discontinuity at the $ 0 $--$\pi$-like transition, as observed in Fig.~\ref{fig:second}\textbf{e}.

It is tempting to explain the reversal of the DC SDE with a similar argument. However, unlike the Josephson inductance which can be probed at any value of $I$, the critical current is, by definition, probed at high current bias. Therefore, if the CPR is close to a local maximum and the current is further increased, the system will jump to its \textit{absolute} maximum~(or minimum). For these reasons, $\Delta I_\mathrm{c}(\lambda_\mathrm{Z})$ is determined by the  absolute maximum and minimum of the CPR when computing $\Delta I_\mathrm{c}(\lambda_\mathrm{Z})$.

The reversal of $\Delta I_\mathrm{c}$  in multi-channel junctions is ultimately due to the complex shape of the CPR, which features \textit{two} maxima and \textit{two} minima at high in-plane fields.
At small $B_y$-fields, both the absolute maximum $I_\mathrm{c}^+$ and the absolute minimum $|I_\mathrm{c}^-|$ decrease with $B_y$, but the latter decrease faster (cf.~Fig.~\ref{fig:first}\textbf{c} and Fig.~\ref{fig:third}\textbf{c}), so that the difference $\Delta I_\mathrm{c}$ increases with nearly constant slope with $B_y$ (cf.~Fig.~\ref{fig:third}\textbf{a,c}). At in-plane field $B_y$ corresponding to Zeeman coupling $\lambda_\mathrm{Z}=1.61$ (gray curve in Fig.~\ref{fig:third}\textbf{e},  below the $ 0 $--$\pi$-like transition), the two minima become degenerate. Eventually, for  higher fields, $I_\mathrm{c}^-$ is determined by the new minimum (the one on the left in Fig.~\ref{fig:third}\textbf{e}, see magenta arrow). The absolute value of this latter minimum decreases much more slowly with $\lambda_\mathrm{Z}$, so that  $|I_\mathrm{c}^-|$ (and consequently $\Delta I_\mathrm{c}$) displays a slope discontinuity at $\lambda_\mathrm{Z}=1.61$, see orange arrow in Fig.~\ref{fig:third}\textbf{c}. For $\lambda_\mathrm{Z}>1.61$,  $I_\mathrm{c}^+$ decreases faster than $|I_\mathrm{c}^-|$, until for $\lambda_\mathrm{Z}=2.14$  the two curves cross, i.e., $\Delta I_\mathrm{c}$ changes sign.
In our model, we observe another slope discontinuity for $\lambda_\mathrm{Z}=2.31$ (purple arrow in Fig.~\ref{fig:third}\textbf{c}), corresponding to the degeneracy of two CPR \textit{maxima}, which induces a negative peak in $\Delta I_\mathrm{c}$, followed by its suppression,
see Fig.~\ref{fig:third}\textbf{d}. 
In the experiment, owing to orbital pair-breaking at high fields (ignored in the model), the visibility of such a negative peak is largely reduced, cf.~Fig.~\ref{fig:third}\textbf{b} and Fig.~\ref{fig:third}\textbf{d}. It is clear from our arguments that the reversal of the supercurrent rectification is not directly related to the $ 0 $--$\pi$-like~transition, although both phenomena originate from the complex character of the CPR in multi-channel junctions.  Our results might explain the often reported observation~\cite{shin2021,Turini2022,Bauriedl2022,Baumgartner2022,BaumgartnerSI2022} of a sharp suppression of the DC SDE above a certain threshold field.

\section{Discussion}

Our findings demonstrate the importance of inductance experiments to extract information that is completely inaccessible in DC~measurements. Unlike critical current measurements, the Josephson inductance can be used to probe the system near equilibrium---for instance, both $L_0$ and $L_0^{\prime}$ can be determined applying an infinitesimally small DC current bias. It is precisely this feature that allows us to reveal the $ 0 $--$ \pi $-like transition. Instead, under large DC bias, the critical current is determined by the absolute maximum (or minimum) of the CPR, and not by relative maxima (or minima).  
For this reason,  $ 0 $--$\pi$-like~transitions can hardly be observed in DC experiments on single junctions.
Our theoretical calculations suggest that the reversal of the AC SDE is much more pronounced than that of the DC SDE, precisely as observed in our experiments, emphasizing the great sensitivity of inductance measurements to probe nonreciprocal supercurrent in Josephson junctions.



In both theory and experiment, we observe only one reversal of the sign of the supercurrent rectification, which is in agreement with, e.g., Refs.~\cite{Turini2022,shin2021}. In contrast, experiments on NiTe$_2$ junctions~\cite{Pal2022} showed multiple oscillations, with $\Delta I_\mathrm{c}$ appearing as a damped sine-function when increasing the in-plane field. In stark contrast, the sudden change of the slope of $\Delta I_\mathrm{c}$ at $B_y=75$~mT in our experiment can certainly not be reproduced by a damped sine.

Finally, we would like to stress that orbital effects, which have recently been invoked to explain supercurrent-rectification effects in ordinary films~\cite{Suri2022,Hou2022}, as well as in epitaxial Al/InAs bilayers~\cite{Sundaresh2022}. These effects do not seem to play any significant role in our experiments, where both the critical current and inductance are completely determined by the Josephson-junction properties only. 


In conclusion, we experimentally demonstrate a reversal of the AC supercurrent diode effect in ballistic Josephson junctions with large Rashba SOI. Based on a minimal theoretical model, we show that this effect provides an experimental signature of $ 0 $--$\pi$-like transitions in the CPR. The accompanying reversal of the critical current nonreciprocity---the DC supercurrent diode effect---is, instead, not directly related to the $ 0 $--$\pi$-like transitions. Nonetheless, both phenomena originate from the distinct spectral properties of the Andreev bound states in 2DEG-based Rashba Josephson junction that give rise to complex Josephson energy landscape, reflecting CPR owing anomalous $\varphi_0$-shifts for different transverse channels and different magnitudes of critical currents for two polarities.

\section{Methods}
\subsection{Experimental details}

Samples are fabricated starting from a semiconductor/metal heterostructure grown by molecular beam epitaxy. The topmost layer (Al, 7~nm) is separated by a quantum well (In$_{0.81}$Ga$_{0.19}$As, 4~nm; InAs, 7~nm; In$_{0.80}$Ga$_{0.20}$As, 10~nm) by two monolayers of GaAs. The quantum well hosts a 2D electron gas with density $n=5\cdot 10^{11}$~cm$^{-2}$ and mobility of approximately $\mu=2.2\cdot 10^{4}$~cm$^{2}$/Vs.
The Josephson junction array is defined on a mesa structure, fabricated via electron beam lithography followed by a wet-etching step. Then, to obtain the weak links, we remove the topmost Al layer by electron beam lithography followed by highly selective Al wet-etching. The selectivity of the etching is crucial in order to obtain ballistic junctions with high transparency. Further details about the sample structure and fabrication are provided in Ref.~\cite{baumgartner2020}.

To measure both the inductance and DC transport characteristics, we embed the sample under study in a circuit featuring a RLC resonator in series with the sample. The circuit  is mounted on the sample holder thermally anchored to the cold finger of a dilution refrigerator. Four leads make possible DC transport measurements in a 4-terminal configuration.  The four leads are connected to the cryostat lines via 1~k\textOmega ~decoupling resistors.   The inductance and capacitance of the RLC resonator are $L_\mathrm{e}=382$~nH and $C_\mathrm{e}=4$~nF, respectively. The resonance frequency for negligible contribution from the sample is, therefore, $f_0=4$~MHz. The inductance of the sample is deduced from the reduction of the resonance frequency with respect to $f_0$. More details on the circuit can be found in Ref.~\cite{baumgartner2020}.

\subsection{Details of the theoretical model}


Our minimal theoretical model allows us to relate the supercurrent of a short ballistic multi-channel Josephson junction with the spectral properties of the associated Andreev bound states. Based on them, we explain the experimentally observed DC and AC SDE.
Assuming that all 2250~junctions within the array are identical, we  focus on one single S-N-S~junction, extract its Josephson CPR, its Josephson energy and thus the Josephson inductance---the inductance of the whole array is just a 2250-multiple of the latter.

We model the short S-N-S~junction as two semi-infinite superconducting~(S) leads that are phase-coherently coupled by a deltalike normal~(N) link~(see~Fig.~1, bottom, in the main text). 
The coherent transport of Cooper pairs through the junction is mediated by Andreev bound states, whose energies and wave functions are obtained by solving the stationary 2D  Bogoliubov--de~Gennes~equation~\cite{DeGennes1989} 
    \begin{equation}
        \left[ \begin{matrix} \hat{\mathcal{H}} & \hat{\Delta} (x) \\ \hat{\Delta}^\dagger (x) & -\hat{\sigma}_y \big( \hat{\mathcal{H}} \big)^* \hat{\sigma}_y \end{matrix} \right] \Psi(x,y) = E \Psi(x,y) , 
        \label{Eq_BdG_1}
    \end{equation}
    where the single-electron Hamiltonian reads as
    \begin{align}
        \hat{\mathcal{H}} &= \hat{\mathcal{H}}_{\mathrm{S},1} \, \Theta(-x) + \hat{\mathcal{H}}_{\mathrm{S},2} \, \Theta(x) + \hat{\mathcal{H}}_\mathrm{N} \, \delta(x) ,
        \intertext{with } 
        \hat{\mathcal{H}}_{\mathrm{S},1/2} &= \left[ -\frac{\hbar^2}{2m} \left( \frac{\partial^2}{\partial x^2} + \frac{\partial^2}{\partial y^2} \right) - \mu \right] \hat{\sigma}_0 + \alpha \left( k_y \hat{\sigma}_x - k_x \hat{\sigma}_y \right) 
        \intertext{and}
        \hat{\mathcal{H}}_\mathrm{N} &= \hat{\mathcal{H}}_{\mathrm{S},1/2} + \left( V_0 \, \hat{\sigma}_0 + V_\mathrm{Z} \, \hat{\sigma}_y \right) d.
    \end{align}

In the above expressions, $\alpha$ parameterizes the Rashba~SOI that is present throughout the whole junction, while $V_{\mathrm{Z}}$ and
$V_0$ stand for the heights of the Zeeman and scalar~(spin-independent) potentials that are spreading over the full width
of the N weak link with effective length $d$. While $V_{\mathrm{Z}}$ takes care about the in-plane magnetic field (aligned along the $ \hat{y} $-axis) that couples to spin via the Zeeman coupling, the reason to introduce $ V_0 $ is merely to capture a reduced transparency of the N~link, caused, e.g., by different electronic densities in the proximitized S and N regions~\cite{Zutic2000}.
Moreover, $ m $ denotes the effective (quasiparticle) mass, $ \mu = \hbar^2 k_\mathrm{F}^2 / (2m) $ the Fermi energy, $ \hat{\sigma}_0 $ the $ 2 \times 2 $ identity matrix, and $ \hat{\sigma}_{x} $, $ \hat{\sigma}_{y} $  stand for the Pauli~matrices.

The $ s $-wave superconducting pairing potential~(induced by the epitaxially grown Al layer) can be written as 
    \begin{equation}
        \hat{\Delta}(x) = \Delta^* \left[ \Theta(-x) + \mathrm{e}^{\mathrm{i} \varphi} \Theta(x) \right] ,
    \end{equation}
with the proximity-induced superconducting gap~$ \Delta^* $ and the superconducting phase difference~$ \varphi $ along the junction. 

To find the Andreev bound states, we (i)~start with the most general ansatz for the in-gap (bound-state) wave functions in the superconductors, (ii)~then eliminate all unknown amplitudes that enter this ansatz by applying appropriate boundary~conditions that should be satisfied at the N~interface, and~(iii)~request the final system of algebraic equations to have a nontrivial solution to get a secular equation. We solve the latter obtaining the Andreev bound state energies, and subsequently also the unknown amplitudes, what gives us the bound-state wave~functions in real space. Having the wave functions, we compute the expectation values of the current operator in the N~region, which is equal to the Josephson current due to charge conservation---so the Cooper pair tunneling is mediated by the available Andreev bound states.
Varying the superconducting phase difference~$ \varphi $, we finally recover the Josephson~CPR~$ I(\varphi) $, the direction-dependent critical~currents~$ I_\mathrm{c}^+ $ and $ I_\mathrm{c}^- $, and the Josephson inductance  

    \begin{equation}
        L(I) = \frac{\hbar}{2e} \frac{\mathrm{d} \varphi(I)}{\mathrm{d} I} ,
    \end{equation}
which we compare with the experimental data; $ e $ refers to the (positive) elementary charge.

To shorten the notation, we define the dimensionless parameters~$ \lambda_\mathrm{SOI} = m \alpha / (\hbar^2 k_\mathrm{F}) $, $ Z = 2 m V_0 d/ (\hbar^2 k_\mathrm{F}) $, and $ \lambda_\mathrm{Z} = 2 m V_\mathrm{Z} d / (\hbar^2 k_\mathrm{F}) $ to quantify the Rashba SOI, the strength of the (scalar) barrier, and the strength of the Zeeman coupling, respectively. 
In agreement with our earlier experimental findings~\cite{baumgartner2020}, we set~$ Z=0.5 $ in all calculations, which corresponds to a junction transparency~\cite{Blonder1982} of~$ \bar{\tau} = 1/[1+(Z/2)^2] \approx 0.94 $, whereas the Rashba SOI~$ \lambda_\mathrm{SOI}=0.661 $ was adapted such that varying the phenomenological Zeeman~parameter~$ \lambda_\mathrm{Z} $ reproduces at best the qualitative magnetic-field dependence of the experimental data.

Although our minimal model includes the Zeeman coupling only in a deltalike manner, we can still exploit the formal analogy with a realistic Zeeman Hamiltonian to convert $ \lambda_\mathrm{Z} $ into a plausible value for the experimental magnetic field. Substituting the formula for the Zeeman gap, $ V_\mathrm{Z} = |g^*| \mu_\mathrm{B} B_y / 2 $, where $ |g^*| \approx 10 $~\cite{Baumgartner2022} and $ \mu_\mathrm{B} $ denotes the Bohr magneton, into the above definition of~$ \lambda_\mathrm{Z} $, the corresponding magnetic field is~$ B_y = \hbar^2 k_\mathrm{F} / (m |g^*| \mu_\mathrm{B} d) \times \lambda_\mathrm{Z} $. For a typical Fermi wave~vector of $ k_\mathrm{F} \approx 3 \times 10^8 \, \mathrm{m}^{-1} $ and N-link length of~$ d = 100 \, \mathrm{nm} $, the $ 0 $--$ \pi $-like transition point at $ \lambda_\mathrm{Z} \approx 2 $ in our model refers then to a magnetic field of~$ B_y \approx 800 \, \mathrm{mT}
$.
This estimate also agrees reasonably well with the  transition field observed in~Ref.~\cite{Dartiailh2021}. What matters to quantitatively map theory and experiment is the elevated ratio of SOI-strength-to-Fermi~level, which in our case 
reads~$ \alpha k_\mathrm{F} / \mu =2 \lambda_\mathrm{SOI} \approx 1.2 $.

In the experiment, the $ 0 $--$ \pi $-like transition occurs already at a substantially smaller magnetic field of about 200~mT, see Fig.~\ref{fig:second}\textbf{c}. 
The reason for this quantitative discrepancy between experiment and theory is most likely the neglect of gap-suppression effects inside the superconducting regions in our theory, as well as the too coarse approximation of the N link by means of a simple delta-function. Extracting the field dependence of the proximity-induced superconducting gap, $ \Delta^*(B_y) $, from the experimental inductance data indeed confirms that the gap becomes substantially suppressed and the Zeeman energy can already overcome the superconducting gap at about~$ 200 \, \mathrm{mT} $. 
As a consequence, the superconducting gap essentially closes for one spin channel, and the junction undergoes the $ 0 $--$ \pi $-like transition.

\begin{acknowledgments}
    Work at Regensburg University was funded by the Deutsche Forschungsgemeinschaft (DFG, German Research Foundation) through Project-ID 314695032–--SFB 1277~(Subprojects B05, B07, and B08)---and Project-ID 454646522---Research grant ``Spin and magnetic properties of superconducting tunnel junctions~(A.C. and J.F.). 
    D.K.~acknowledges partial support from the project 
IM-2021-26 (SUPERSPIN) funded by the Slovak Academy of Sciences via the programme IMPULZ 2021.
\end{acknowledgments}
\vspace{2mm}

 \noindent {\it Author Contributions}: 
            $^\dagger$A.~Costa and $^\dagger$C.~Baumgartner   contributed equally to this work. C.B.~ and J.B.~fabricated the devices and performed the measurements. S.R.~developed and optimized the measurement method. 
            T.L., S.G., and G.C.G.~designed the heterostructure, conducted MBE growth and performed initial characterization of the hybrid superconductor/semiconductor wafer. C.B.~and N.P.~analyzed the data. N.P.~and C.S.~conceived the experiment. A.C., D.K.,~and J.F.~formulated the theoretical model.    C.S.~and M.J.M.~supervised research activities at Regensburg and Purdue, respectively. N.P., A.C.,~and D.K.~wrote the manuscript. All authors contributed to discussions and to the writing of the manuscript.


\begin{thebibliography}{61}%
\makeatletter
\providecommand \@ifxundefined [1]{%
 \@ifx{#1\undefined}
}%
\providecommand \@ifnum [1]{%
 \ifnum #1\expandafter \@firstoftwo
 \else \expandafter \@secondoftwo
 \fi
}%
\providecommand \@ifx [1]{%
 \ifx #1\expandafter \@firstoftwo
 \else \expandafter \@secondoftwo
 \fi
}%
\providecommand \natexlab [1]{#1}%
\providecommand \enquote  [1]{``#1''}%
\providecommand \bibnamefont  [1]{#1}%
\providecommand \bibfnamefont [1]{#1}%
\providecommand \citenamefont [1]{#1}%
\providecommand \href@noop [0]{\@secondoftwo}%
\providecommand \href [0]{\begingroup \@sanitize@url \@href}%
\providecommand \@href[1]{\@@startlink{#1}\@@href}%
\providecommand \@@href[1]{\endgroup#1\@@endlink}%
\providecommand \@sanitize@url [0]{\catcode `\\12\catcode `\$12\catcode
  `\&12\catcode `\#12\catcode `\^12\catcode `\_12\catcode `\%12\relax}%
\providecommand \@@startlink[1]{}%
\providecommand \@@endlink[0]{}%
\providecommand \url  [0]{\begingroup\@sanitize@url \@url }%
\providecommand \@url [1]{\endgroup\@href {#1}{\urlprefix }}%
\providecommand \urlprefix  [0]{URL }%
\providecommand \Eprint [0]{\href }%
\providecommand \doibase [0]{https://doi.org/}%
\providecommand \selectlanguage [0]{\@gobble}%
\providecommand \bibinfo  [0]{\@secondoftwo}%
\providecommand \bibfield  [0]{\@secondoftwo}%
\providecommand \translation [1]{[#1]}%
\providecommand \BibitemOpen [0]{}%
\providecommand \bibitemStop [0]{}%
\providecommand \bibitemNoStop [0]{.\EOS\space}%
\providecommand \EOS [0]{\spacefactor3000\relax}%
\providecommand \BibitemShut  [1]{\csname bibitem#1\endcsname}%
\let\auto@bib@innerbib\@empty
\bibitem [{\citenamefont {Ando}\ \emph {et~al.}(2020)\citenamefont {Ando},
  \citenamefont {Miyasaka}, \citenamefont {Li}, \citenamefont {Ishizuka},
  \citenamefont {Arakawa}, \citenamefont {Shiota}, \citenamefont {Moriyama},
  \citenamefont {Yanase},\ and\ \citenamefont {Ono}}]{Ando2020}%
  \BibitemOpen
  \bibfield  {author} {\bibinfo {author} {\bibfnamefont {F.}~\bibnamefont
  {Ando}}, \bibinfo {author} {\bibfnamefont {Y.}~\bibnamefont {Miyasaka}},
  \bibinfo {author} {\bibfnamefont {T.}~\bibnamefont {Li}}, \bibinfo {author}
  {\bibfnamefont {J.}~\bibnamefont {Ishizuka}}, \bibinfo {author}
  {\bibfnamefont {T.}~\bibnamefont {Arakawa}}, \bibinfo {author} {\bibfnamefont
  {Y.}~\bibnamefont {Shiota}}, \bibinfo {author} {\bibfnamefont
  {T.}~\bibnamefont {Moriyama}}, \bibinfo {author} {\bibfnamefont
  {Y.}~\bibnamefont {Yanase}},\ and\ \bibinfo {author} {\bibfnamefont
  {T.}~\bibnamefont {Ono}},\ }\bibfield  {title} {\bibinfo {title} {Observation
  of superconducting diode effect},\ }\href
  {https://doi.org/10.1038/s41586-020-2590-4} {\bibfield  {journal} {\bibinfo
  {journal} {Nature}\ }\textbf {\bibinfo {volume} {584}},\ \bibinfo {pages}
  {373} (\bibinfo {year} {2020})}\BibitemShut {NoStop}%
\bibitem [{\citenamefont {Baumgartner}\ \emph
  {et~al.}(2022{\natexlab{a}})\citenamefont {Baumgartner}, \citenamefont
  {Fuchs}, \citenamefont {Costa}, \citenamefont {Reinhardt}, \citenamefont
  {Gronin}, \citenamefont {Gardner}, \citenamefont {Lindemann}, \citenamefont
  {Manfra}, \citenamefont {Faria~Junior}, \citenamefont {Kochan}, \citenamefont
  {Fabian}, \citenamefont {Paradiso},\ and\ \citenamefont
  {Strunk}}]{Baumgartner2022}%
  \BibitemOpen
  \bibfield  {author} {\bibinfo {author} {\bibfnamefont {C.}~\bibnamefont
  {Baumgartner}}, \bibinfo {author} {\bibfnamefont {L.}~\bibnamefont {Fuchs}},
  \bibinfo {author} {\bibfnamefont {A.}~\bibnamefont {Costa}}, \bibinfo
  {author} {\bibfnamefont {S.}~\bibnamefont {Reinhardt}}, \bibinfo {author}
  {\bibfnamefont {S.}~\bibnamefont {Gronin}}, \bibinfo {author} {\bibfnamefont
  {G.~C.}\ \bibnamefont {Gardner}}, \bibinfo {author} {\bibfnamefont
  {T.}~\bibnamefont {Lindemann}}, \bibinfo {author} {\bibfnamefont {M.~J.}\
  \bibnamefont {Manfra}}, \bibinfo {author} {\bibfnamefont {P.~E.}\
  \bibnamefont {Faria~Junior}}, \bibinfo {author} {\bibfnamefont
  {D.}~\bibnamefont {Kochan}}, \bibinfo {author} {\bibfnamefont
  {J.}~\bibnamefont {Fabian}}, \bibinfo {author} {\bibfnamefont
  {N.}~\bibnamefont {Paradiso}},\ and\ \bibinfo {author} {\bibfnamefont
  {C.}~\bibnamefont {Strunk}},\ }\bibfield  {title} {\bibinfo {title}
  {{Supercurrent rectification and magnetochiral effects in symmetric Josephson
  junctions}},\ }\href {https://doi.org/10.1038/s41565-021-01009-9} {\bibfield
  {journal} {\bibinfo  {journal} {Nature Nanotechnology}\ }\textbf {\bibinfo
  {volume} {17}},\ \bibinfo {pages} {39} (\bibinfo {year}
  {2022}{\natexlab{a}})}\BibitemShut {NoStop}%
\bibitem [{\citenamefont {Baumgartner}\ \emph
  {et~al.}(2022{\natexlab{b}})\citenamefont {Baumgartner}, \citenamefont
  {Fuchs}, \citenamefont {Costa}, \citenamefont {Pic{\'{o}}-Cort{\'{e}}s},
  \citenamefont {Reinhardt}, \citenamefont {Gronin}, \citenamefont {Gardner},
  \citenamefont {Lindemann}, \citenamefont {Manfra}, \citenamefont {Junior},
  \citenamefont {Kochan}, \citenamefont {Fabian}, \citenamefont {Paradiso},\
  and\ \citenamefont {Strunk}}]{BaumgartnerSI2022}%
  \BibitemOpen
  \bibfield  {author} {\bibinfo {author} {\bibfnamefont {C.}~\bibnamefont
  {Baumgartner}}, \bibinfo {author} {\bibfnamefont {L.}~\bibnamefont {Fuchs}},
  \bibinfo {author} {\bibfnamefont {A.}~\bibnamefont {Costa}}, \bibinfo
  {author} {\bibfnamefont {J.}~\bibnamefont {Pic{\'{o}}-Cort{\'{e}}s}},
  \bibinfo {author} {\bibfnamefont {S.}~\bibnamefont {Reinhardt}}, \bibinfo
  {author} {\bibfnamefont {S.}~\bibnamefont {Gronin}}, \bibinfo {author}
  {\bibfnamefont {G.~C.}\ \bibnamefont {Gardner}}, \bibinfo {author}
  {\bibfnamefont {T.}~\bibnamefont {Lindemann}}, \bibinfo {author}
  {\bibfnamefont {M.~J.}\ \bibnamefont {Manfra}}, \bibinfo {author}
  {\bibfnamefont {P.~E.~F.}\ \bibnamefont {Junior}}, \bibinfo {author}
  {\bibfnamefont {D.}~\bibnamefont {Kochan}}, \bibinfo {author} {\bibfnamefont
  {J.}~\bibnamefont {Fabian}}, \bibinfo {author} {\bibfnamefont
  {N.}~\bibnamefont {Paradiso}},\ and\ \bibinfo {author} {\bibfnamefont
  {C.}~\bibnamefont {Strunk}},\ }\bibfield  {title} {\bibinfo {title} {{Effect
  of Rashba and Dresselhaus spin{\textendash}orbit coupling on supercurrent
  rectification and magnetochiral anisotropy of ballistic Josephson
  junctions}},\ }\href {https://doi.org/10.1088/1361-648x/ac4d5e} {\bibfield
  {journal} {\bibinfo  {journal} {Journal of Physics: Condensed Matter}\
  }\textbf {\bibinfo {volume} {34}},\ \bibinfo {pages} {154005} (\bibinfo
  {year} {2022}{\natexlab{b}})}\BibitemShut {NoStop}%
\bibitem [{\citenamefont {Wu}\ \emph {et~al.}(2022)\citenamefont {Wu},
  \citenamefont {Wang}, \citenamefont {Xu}, \citenamefont {Sivakumar},
  \citenamefont {Pasco}, \citenamefont {Filippozzi}, \citenamefont {Parkin},
  \citenamefont {Zeng}, \citenamefont {McQueen},\ and\ \citenamefont
  {Ali}}]{Wu2022}%
  \BibitemOpen
  \bibfield  {author} {\bibinfo {author} {\bibfnamefont {H.}~\bibnamefont
  {Wu}}, \bibinfo {author} {\bibfnamefont {Y.}~\bibnamefont {Wang}}, \bibinfo
  {author} {\bibfnamefont {Y.}~\bibnamefont {Xu}}, \bibinfo {author}
  {\bibfnamefont {P.~K.}\ \bibnamefont {Sivakumar}}, \bibinfo {author}
  {\bibfnamefont {C.}~\bibnamefont {Pasco}}, \bibinfo {author} {\bibfnamefont
  {U.}~\bibnamefont {Filippozzi}}, \bibinfo {author} {\bibfnamefont {S.~S.~P.}\
  \bibnamefont {Parkin}}, \bibinfo {author} {\bibfnamefont {Y.-J.}\
  \bibnamefont {Zeng}}, \bibinfo {author} {\bibfnamefont {T.}~\bibnamefont
  {McQueen}},\ and\ \bibinfo {author} {\bibfnamefont {M.~N.}\ \bibnamefont
  {Ali}},\ }\bibfield  {title} {\bibinfo {title} {{The field-free Josephson
  diode in a van der Waals heterostructure}},\ }\href
  {https://doi.org/10.1038/s41586-022-04504-8} {\bibfield  {journal} {\bibinfo
  {journal} {Nature}\ }\textbf {\bibinfo {volume} {604}},\ \bibinfo {pages}
  {653} (\bibinfo {year} {2022})}\BibitemShut {NoStop}%
\bibitem [{\citenamefont {Jeon}\ \emph {et~al.}(2022)\citenamefont {Jeon},
  \citenamefont {Kim}, \citenamefont {Yoon}, \citenamefont {Jeon},
  \citenamefont {Han}, \citenamefont {Cottet}, \citenamefont {Kontos},\ and\
  \citenamefont {Parkin}}]{Jeon2022}%
  \BibitemOpen
  \bibfield  {author} {\bibinfo {author} {\bibfnamefont {K.-R.}\ \bibnamefont
  {Jeon}}, \bibinfo {author} {\bibfnamefont {J.-K.}\ \bibnamefont {Kim}},
  \bibinfo {author} {\bibfnamefont {J.}~\bibnamefont {Yoon}}, \bibinfo {author}
  {\bibfnamefont {J.-C.}\ \bibnamefont {Jeon}}, \bibinfo {author}
  {\bibfnamefont {H.}~\bibnamefont {Han}}, \bibinfo {author} {\bibfnamefont
  {A.}~\bibnamefont {Cottet}}, \bibinfo {author} {\bibfnamefont
  {T.}~\bibnamefont {Kontos}},\ and\ \bibinfo {author} {\bibfnamefont
  {S.~S.~P.}\ \bibnamefont {Parkin}},\ }\bibfield  {title} {\bibinfo {title}
  {{Zero-field polarity-reversible Josephson supercurrent diodes enabled by a
  proximity-magnetized Pt barrier}},\ }\href
  {https://doi.org/10.1038/s41563-022-01300-7} {\bibfield  {journal} {\bibinfo
  {journal} {Nature Materials}\ }\textbf {\bibinfo {volume} {21}},\ \bibinfo
  {pages} {1008} (\bibinfo {year} {2022})}\BibitemShut {NoStop}%
\bibitem [{\citenamefont {Pal}\ \emph {et~al.}(2022)\citenamefont {Pal},
  \citenamefont {Chakraborty}, \citenamefont {Sivakumar}, \citenamefont
  {Davydova}, \citenamefont {Gopi}, \citenamefont {Pandeya}, \citenamefont
  {Krieger}, \citenamefont {Zhang}, \citenamefont {Date}, \citenamefont {Ju},
  \citenamefont {Yuan}, \citenamefont {Schr{\"o}ter}, \citenamefont {Fu},\ and\
  \citenamefont {Parkin}}]{Pal2022}%
  \BibitemOpen
  \bibfield  {author} {\bibinfo {author} {\bibfnamefont {B.}~\bibnamefont
  {Pal}}, \bibinfo {author} {\bibfnamefont {A.}~\bibnamefont {Chakraborty}},
  \bibinfo {author} {\bibfnamefont {P.~K.}\ \bibnamefont {Sivakumar}}, \bibinfo
  {author} {\bibfnamefont {M.}~\bibnamefont {Davydova}}, \bibinfo {author}
  {\bibfnamefont {A.~K.}\ \bibnamefont {Gopi}}, \bibinfo {author}
  {\bibfnamefont {A.~K.}\ \bibnamefont {Pandeya}}, \bibinfo {author}
  {\bibfnamefont {J.~A.}\ \bibnamefont {Krieger}}, \bibinfo {author}
  {\bibfnamefont {Y.}~\bibnamefont {Zhang}}, \bibinfo {author} {\bibfnamefont
  {M.}~\bibnamefont {Date}}, \bibinfo {author} {\bibfnamefont {S.}~\bibnamefont
  {Ju}}, \bibinfo {author} {\bibfnamefont {N.}~\bibnamefont {Yuan}}, \bibinfo
  {author} {\bibfnamefont {N.~B.~M.}\ \bibnamefont {Schr{\"o}ter}}, \bibinfo
  {author} {\bibfnamefont {L.}~\bibnamefont {Fu}},\ and\ \bibinfo {author}
  {\bibfnamefont {S.~S.~P.}\ \bibnamefont {Parkin}},\ }\bibfield  {title}
  {\bibinfo {title} {{Josephson diode effect from Cooper pair momentum in a
  topological semimetal}},\ }\bibfield  {journal} {\bibinfo  {journal} {Nature
  Physics}\ }\href {https://doi.org/10.1038/s41567-022-01699-5}
  {10.1038/s41567-022-01699-5} (\bibinfo {year} {2022})\BibitemShut {NoStop}%
\bibitem [{\citenamefont {Bauriedl}\ \emph {et~al.}(2022)\citenamefont
  {Bauriedl}, \citenamefont {B{\"a}uml}, \citenamefont {Fuchs}, \citenamefont
  {Baumgartner}, \citenamefont {Paulik}, \citenamefont {Bauer}, \citenamefont
  {Lin}, \citenamefont {Lupton}, \citenamefont {Taniguchi}, \citenamefont
  {Watanabe}, \citenamefont {Strunk},\ and\ \citenamefont
  {Paradiso}}]{Bauriedl2022}%
  \BibitemOpen
  \bibfield  {author} {\bibinfo {author} {\bibfnamefont {L.}~\bibnamefont
  {Bauriedl}}, \bibinfo {author} {\bibfnamefont {C.}~\bibnamefont {B{\"a}uml}},
  \bibinfo {author} {\bibfnamefont {L.}~\bibnamefont {Fuchs}}, \bibinfo
  {author} {\bibfnamefont {C.}~\bibnamefont {Baumgartner}}, \bibinfo {author}
  {\bibfnamefont {N.}~\bibnamefont {Paulik}}, \bibinfo {author} {\bibfnamefont
  {J.~M.}\ \bibnamefont {Bauer}}, \bibinfo {author} {\bibfnamefont {K.-Q.}\
  \bibnamefont {Lin}}, \bibinfo {author} {\bibfnamefont {J.~M.}\ \bibnamefont
  {Lupton}}, \bibinfo {author} {\bibfnamefont {T.}~\bibnamefont {Taniguchi}},
  \bibinfo {author} {\bibfnamefont {K.}~\bibnamefont {Watanabe}}, \bibinfo
  {author} {\bibfnamefont {C.}~\bibnamefont {Strunk}},\ and\ \bibinfo {author}
  {\bibfnamefont {N.}~\bibnamefont {Paradiso}},\ }\bibfield  {title} {\bibinfo
  {title} {{Supercurrent diode effect and magnetochiral anisotropy in few-layer
  NbSe2}},\ }\href {https://doi.org/10.1038/s41467-022-31954-5} {\bibfield
  {journal} {\bibinfo  {journal} {Nature Communications}\ }\textbf {\bibinfo
  {volume} {13}},\ \bibinfo {pages} {4266} (\bibinfo {year}
  {2022})}\BibitemShut {NoStop}%
\bibitem [{\citenamefont {Turini}\ \emph {et~al.}(2022)\citenamefont {Turini},
  \citenamefont {Salimian}, \citenamefont {Carrega}, \citenamefont {Iorio},
  \citenamefont {Strambini}, \citenamefont {Giazotto}, \citenamefont {Zannier},
  \citenamefont {Sorba},\ and\ \citenamefont {Heun}}]{Turini2022}%
  \BibitemOpen
  \bibfield  {author} {\bibinfo {author} {\bibfnamefont {B.}~\bibnamefont
  {Turini}}, \bibinfo {author} {\bibfnamefont {S.}~\bibnamefont {Salimian}},
  \bibinfo {author} {\bibfnamefont {M.}~\bibnamefont {Carrega}}, \bibinfo
  {author} {\bibfnamefont {A.}~\bibnamefont {Iorio}}, \bibinfo {author}
  {\bibfnamefont {E.}~\bibnamefont {Strambini}}, \bibinfo {author}
  {\bibfnamefont {F.}~\bibnamefont {Giazotto}}, \bibinfo {author}
  {\bibfnamefont {V.}~\bibnamefont {Zannier}}, \bibinfo {author} {\bibfnamefont
  {L.}~\bibnamefont {Sorba}},\ and\ \bibinfo {author} {\bibfnamefont
  {S.}~\bibnamefont {Heun}},\ }\href
  {https://doi.org/10.48550/ARXIV.2207.08772} {\bibinfo {title} {{Josephson
  Diode Effect in High Mobility InSb Nanoflags}}} (\bibinfo {year}
  {2022})\BibitemShut {NoStop}%
\bibitem [{\citenamefont {Zhang}\ \emph {et~al.}(2022)\citenamefont {Zhang},
  \citenamefont {Li}, \citenamefont {Aguilar}, \citenamefont {Zhang},
  \citenamefont {Pendharkar}, \citenamefont {Dempsey}, \citenamefont {Lee},
  \citenamefont {Harrington}, \citenamefont {Tan}, \citenamefont {Meyer},
  \citenamefont {Houzet}, \citenamefont {Palmstrom},\ and\ \citenamefont
  {Frolov}}]{FrolovQW2022}%
  \BibitemOpen
  \bibfield  {author} {\bibinfo {author} {\bibfnamefont {B.}~\bibnamefont
  {Zhang}}, \bibinfo {author} {\bibfnamefont {Z.}~\bibnamefont {Li}}, \bibinfo
  {author} {\bibfnamefont {V.}~\bibnamefont {Aguilar}}, \bibinfo {author}
  {\bibfnamefont {P.}~\bibnamefont {Zhang}}, \bibinfo {author} {\bibfnamefont
  {M.}~\bibnamefont {Pendharkar}}, \bibinfo {author} {\bibfnamefont
  {C.}~\bibnamefont {Dempsey}}, \bibinfo {author} {\bibfnamefont {J.~S.}\
  \bibnamefont {Lee}}, \bibinfo {author} {\bibfnamefont {S.~D.}\ \bibnamefont
  {Harrington}}, \bibinfo {author} {\bibfnamefont {S.}~\bibnamefont {Tan}},
  \bibinfo {author} {\bibfnamefont {J.~S.}\ \bibnamefont {Meyer}}, \bibinfo
  {author} {\bibfnamefont {M.}~\bibnamefont {Houzet}}, \bibinfo {author}
  {\bibfnamefont {C.~J.}\ \bibnamefont {Palmstrom}},\ and\ \bibinfo {author}
  {\bibfnamefont {S.~M.}\ \bibnamefont {Frolov}},\ }\href@noop {} {\bibinfo
  {title} {{Evidence of $\varphi_0$-Josephson junction from skewed diffraction
  patterns in Sn-InSb nanowires}}} (\bibinfo {year} {2022}),\ \Eprint
  {https://arxiv.org/abs/2212.00199} {arXiv:2212.00199 [cond-mat.supr-con]}
  \BibitemShut {NoStop}%
\bibitem [{\citenamefont {Diez-Merida}\ \emph {et~al.}(2021)\citenamefont
  {Diez-Merida}, \citenamefont {Diez-Carlon}, \citenamefont {Yang},
  \citenamefont {Xie}, \citenamefont {Gao}, \citenamefont {Watanabe},
  \citenamefont {Taniguchi}, \citenamefont {Lu}, \citenamefont {Law},\ and\
  \citenamefont {Efetov}}]{diezmerida2021magnetic}%
  \BibitemOpen
  \bibfield  {author} {\bibinfo {author} {\bibfnamefont {J.}~\bibnamefont
  {Diez-Merida}}, \bibinfo {author} {\bibfnamefont {A.}~\bibnamefont
  {Diez-Carlon}}, \bibinfo {author} {\bibfnamefont {S.~Y.}\ \bibnamefont
  {Yang}}, \bibinfo {author} {\bibfnamefont {Y.~M.}\ \bibnamefont {Xie}},
  \bibinfo {author} {\bibfnamefont {X.~J.}\ \bibnamefont {Gao}}, \bibinfo
  {author} {\bibfnamefont {K.}~\bibnamefont {Watanabe}}, \bibinfo {author}
  {\bibfnamefont {T.}~\bibnamefont {Taniguchi}}, \bibinfo {author}
  {\bibfnamefont {X.}~\bibnamefont {Lu}}, \bibinfo {author} {\bibfnamefont
  {K.~T.}\ \bibnamefont {Law}},\ and\ \bibinfo {author} {\bibfnamefont {D.~K.}\
  \bibnamefont {Efetov}},\ }\href@noop {} {\bibinfo {title} {{Magnetic
  Josephson Junctions and Superconducting Diodes in Magic Angle Twisted Bilayer
  Graphene}}} (\bibinfo {year} {2021}),\ \Eprint
  {https://arxiv.org/abs/2110.01067} {arXiv:2110.01067 [cond-mat.supr-con]}
  \BibitemShut {NoStop}%
\bibitem [{\citenamefont {Lin}\ \emph {et~al.}(2022)\citenamefont {Lin},
  \citenamefont {Siriviboon}, \citenamefont {Scammell}, \citenamefont {Liu},
  \citenamefont {Rhodes}, \citenamefont {Watanabe}, \citenamefont {Taniguchi},
  \citenamefont {Hone}, \citenamefont {Scheurer},\ and\ \citenamefont
  {Li}}]{lin2022zerofield}%
  \BibitemOpen
  \bibfield  {author} {\bibinfo {author} {\bibfnamefont {J.-X.}\ \bibnamefont
  {Lin}}, \bibinfo {author} {\bibfnamefont {P.}~\bibnamefont {Siriviboon}},
  \bibinfo {author} {\bibfnamefont {H.~D.}\ \bibnamefont {Scammell}}, \bibinfo
  {author} {\bibfnamefont {S.}~\bibnamefont {Liu}}, \bibinfo {author}
  {\bibfnamefont {D.}~\bibnamefont {Rhodes}}, \bibinfo {author} {\bibfnamefont
  {K.}~\bibnamefont {Watanabe}}, \bibinfo {author} {\bibfnamefont
  {T.}~\bibnamefont {Taniguchi}}, \bibinfo {author} {\bibfnamefont
  {J.}~\bibnamefont {Hone}}, \bibinfo {author} {\bibfnamefont {M.~S.}\
  \bibnamefont {Scheurer}},\ and\ \bibinfo {author} {\bibfnamefont {J.~I.~A.}\
  \bibnamefont {Li}},\ }\bibfield  {title} {\bibinfo {title} {Zero-field
  superconducting diode effect in small-twist-angle trilayer graphene},\ }\href
  {https://doi.org/10.1038/s41567-022-01700-1} {\bibfield  {journal} {\bibinfo
  {journal} {Nature Physics}\ }\textbf {\bibinfo {volume} {18}},\ \bibinfo
  {pages} {1221} (\bibinfo {year} {2022})}\BibitemShut {NoStop}%
\bibitem [{\citenamefont {Scammell}\ \emph {et~al.}(2022)\citenamefont
  {Scammell}, \citenamefont {Li},\ and\ \citenamefont
  {Scheurer}}]{scammell2022theory}%
  \BibitemOpen
  \bibfield  {author} {\bibinfo {author} {\bibfnamefont {H.~D.}\ \bibnamefont
  {Scammell}}, \bibinfo {author} {\bibfnamefont {J.~I.~A.}\ \bibnamefont
  {Li}},\ and\ \bibinfo {author} {\bibfnamefont {M.~S.}\ \bibnamefont
  {Scheurer}},\ }\bibfield  {title} {\bibinfo {title} {Theory of zero-field
  superconducting diode effect in twisted trilayer graphene},\ }\href
  {https://doi.org/10.1088/2053-1583/ac5b16} {\bibfield  {journal} {\bibinfo
  {journal} {2D Materials}\ }\textbf {\bibinfo {volume} {9}},\ \bibinfo {pages}
  {025027} (\bibinfo {year} {2022})}\BibitemShut {NoStop}%
\bibitem [{\citenamefont {Lu}\ \emph {et~al.}(2022)\citenamefont {Lu},
  \citenamefont {Ikegaya}, \citenamefont {Burset}, \citenamefont {Tanaka},\
  and\ \citenamefont {Nagaosa}}]{BoLu2022}%
  \BibitemOpen
  \bibfield  {author} {\bibinfo {author} {\bibfnamefont {B.}~\bibnamefont
  {Lu}}, \bibinfo {author} {\bibfnamefont {S.}~\bibnamefont {Ikegaya}},
  \bibinfo {author} {\bibfnamefont {P.}~\bibnamefont {Burset}}, \bibinfo
  {author} {\bibfnamefont {Y.}~\bibnamefont {Tanaka}},\ and\ \bibinfo {author}
  {\bibfnamefont {N.}~\bibnamefont {Nagaosa}},\ }\href@noop {} {\bibinfo
  {title} {Josephson diode effect on the surface of topological insulators}}
  (\bibinfo {year} {2022}),\ \Eprint {https://arxiv.org/abs/2211.10572}
  {arXiv:2211.10572 [cond-mat.supr-con]} \BibitemShut {NoStop}%
\bibitem [{\citenamefont {Fu}\ \emph {et~al.}(2022)\citenamefont {Fu},
  \citenamefont {Xu}, \citenamefont {Lee}, \citenamefont {Ang},\ and\
  \citenamefont {Liu}}]{Fu2022TopoDiode}%
  \BibitemOpen
  \bibfield  {author} {\bibinfo {author} {\bibfnamefont {P.-H.}\ \bibnamefont
  {Fu}}, \bibinfo {author} {\bibfnamefont {Y.}~\bibnamefont {Xu}}, \bibinfo
  {author} {\bibfnamefont {C.~H.}\ \bibnamefont {Lee}}, \bibinfo {author}
  {\bibfnamefont {Y.~S.}\ \bibnamefont {Ang}},\ and\ \bibinfo {author}
  {\bibfnamefont {J.-F.}\ \bibnamefont {Liu}},\ }\href@noop {} {\bibinfo
  {title} {{Gate-Tunable High-Efficiency Topological Josephson Diode}}}
  (\bibinfo {year} {2022}),\ \Eprint {https://arxiv.org/abs/2212.01980}
  {arXiv:2212.01980 [cond-mat.supr-con]} \BibitemShut {NoStop}%
\bibitem [{\citenamefont {Daido}\ \emph {et~al.}(2022)\citenamefont {Daido},
  \citenamefont {Ikeda},\ and\ \citenamefont {Yanase}}]{daido2021prl}%
  \BibitemOpen
  \bibfield  {author} {\bibinfo {author} {\bibfnamefont {A.}~\bibnamefont
  {Daido}}, \bibinfo {author} {\bibfnamefont {Y.}~\bibnamefont {Ikeda}},\ and\
  \bibinfo {author} {\bibfnamefont {Y.}~\bibnamefont {Yanase}},\ }\bibfield
  {title} {\bibinfo {title} {Intrinsic superconducting diode effect},\ }\href
  {https://doi.org/10.1103/PhysRevLett.128.037001} {\bibfield  {journal}
  {\bibinfo  {journal} {Phys. Rev. Lett.}\ }\textbf {\bibinfo {volume} {128}},\
  \bibinfo {pages} {037001} (\bibinfo {year} {2022})}\BibitemShut {NoStop}%
\bibitem [{\citenamefont {Yuan}\ and\ \citenamefont {Fu}(2022)}]{yuan2021pnas}%
  \BibitemOpen
  \bibfield  {author} {\bibinfo {author} {\bibfnamefont {N.~F.~Q.}\
  \bibnamefont {Yuan}}\ and\ \bibinfo {author} {\bibfnamefont {L.}~\bibnamefont
  {Fu}},\ }\bibfield  {title} {\bibinfo {title} {{Supercurrent diode effect and
  finite-momentum superconductors}},\ }\href
  {https://doi.org/10.1073/pnas.2119548119} {\bibfield  {journal} {\bibinfo
  {journal} {Proceedings of the National Academy of Sciences}\ }\textbf
  {\bibinfo {volume} {119}},\ \bibinfo {pages} {e2119548119} (\bibinfo {year}
  {2022})},\ \Eprint
  {https://arxiv.org/abs/https://www.pnas.org/doi/pdf/10.1073/pnas.2119548119}
  {https://www.pnas.org/doi/pdf/10.1073/pnas.2119548119} \BibitemShut {NoStop}%
\bibitem [{\citenamefont {He}\ \emph {et~al.}(2022)\citenamefont {He},
  \citenamefont {Tanaka},\ and\ \citenamefont {Nagaosa}}]{he2021njp}%
  \BibitemOpen
  \bibfield  {author} {\bibinfo {author} {\bibfnamefont {J.~J.}\ \bibnamefont
  {He}}, \bibinfo {author} {\bibfnamefont {Y.}~\bibnamefont {Tanaka}},\ and\
  \bibinfo {author} {\bibfnamefont {N.}~\bibnamefont {Nagaosa}},\ }\bibfield
  {title} {\bibinfo {title} {A phenomenological theory of superconductor
  diodes},\ }\href {https://doi.org/10.1088/1367-2630/ac6766} {\bibfield
  {journal} {\bibinfo  {journal} {New Journal of Physics}\ }\textbf {\bibinfo
  {volume} {24}},\ \bibinfo {pages} {053014} (\bibinfo {year}
  {2022})}\BibitemShut {NoStop}%
\bibitem [{\citenamefont {Ili\ifmmode~\acute{c}\else \'{c}\fi{}}\ and\
  \citenamefont {Bergeret}(2022)}]{Ilic2022}%
  \BibitemOpen
  \bibfield  {author} {\bibinfo {author} {\bibfnamefont {S.}~\bibnamefont
  {Ili\ifmmode~\acute{c}\else \'{c}\fi{}}}\ and\ \bibinfo {author}
  {\bibfnamefont {F.~S.}\ \bibnamefont {Bergeret}},\ }\bibfield  {title}
  {\bibinfo {title} {{Theory of the Supercurrent Diode Effect in Rashba
  Superconductors with Arbitrary Disorder}},\ }\href
  {https://doi.org/10.1103/PhysRevLett.128.177001} {\bibfield  {journal}
  {\bibinfo  {journal} {Phys. Rev. Lett.}\ }\textbf {\bibinfo {volume} {128}},\
  \bibinfo {pages} {177001} (\bibinfo {year} {2022})}\BibitemShut {NoStop}%
\bibitem [{\citenamefont {Andreev}(1966)}]{Andreev1966}%
  \BibitemOpen
  \bibfield  {author} {\bibinfo {author} {\bibfnamefont {A.~F.}\ \bibnamefont
  {Andreev}},\ }\bibfield  {title} {\bibinfo {title} {{Electron Spectrum of the
  Intermediate State of Superconductors}},\ }\href@noop {} {\bibfield
  {journal} {\bibinfo  {journal} {Zh. Eksp. Teor. Fiz.}\ }\textbf {\bibinfo
  {volume} {49}},\ \bibinfo {pages} {655} (\bibinfo {year} {1966})}\BibitemShut
  {NoStop}%
\bibitem [{And(1966)}]{Andreev1966alt}%
  \BibitemOpen
  \href {http://www.jetp.ac.ru/cgi-bin/e/index/e/22/2/p455?a=list} {\bibfield
  {journal} {\bibinfo  {journal} {J. Exp. Theor. Phys.}\ }\textbf {\bibinfo
  {volume} {22}},\ \bibinfo {pages} {455} (\bibinfo {year} {1966})}\BibitemShut
  {NoStop}%
\bibitem [{\citenamefont {Bezuglyi}\ \emph {et~al.}(2002)\citenamefont
  {Bezuglyi}, \citenamefont {Rozhavsky}, \citenamefont {Vagner},\ and\
  \citenamefont {Wyder}}]{Bezuglyi2002}%
  \BibitemOpen
  \bibfield  {author} {\bibinfo {author} {\bibfnamefont {E.~V.}\ \bibnamefont
  {Bezuglyi}}, \bibinfo {author} {\bibfnamefont {A.~S.}\ \bibnamefont
  {Rozhavsky}}, \bibinfo {author} {\bibfnamefont {I.~D.}\ \bibnamefont
  {Vagner}},\ and\ \bibinfo {author} {\bibfnamefont {P.}~\bibnamefont
  {Wyder}},\ }\bibfield  {title} {\bibinfo {title} {{Combined effect of Zeeman
  splitting and spin-orbit interaction on the Josephson current in a
  superconductor--two-dimensional electron gas--superconductor structure}},\
  }\href {https://doi.org/10.1103/PhysRevB.66.052508} {\bibfield  {journal}
  {\bibinfo  {journal} {Phys. Rev. B}\ }\textbf {\bibinfo {volume} {66}},\
  \bibinfo {pages} {052508} (\bibinfo {year} {2002})}\BibitemShut {NoStop}%
\bibitem [{\citenamefont {Krive}\ \emph {et~al.}(2004)\citenamefont {Krive},
  \citenamefont {Gorelik}, \citenamefont {Shekhter},\ and\ \citenamefont
  {Jonson}}]{Krive2004}%
  \BibitemOpen
  \bibfield  {author} {\bibinfo {author} {\bibfnamefont {I.~V.}\ \bibnamefont
  {Krive}}, \bibinfo {author} {\bibfnamefont {L.~Y.}\ \bibnamefont {Gorelik}},
  \bibinfo {author} {\bibfnamefont {R.~I.}\ \bibnamefont {Shekhter}},\ and\
  \bibinfo {author} {\bibfnamefont {M.}~\bibnamefont {Jonson}},\ }\bibfield
  {title} {\bibinfo {title} {{Chiral symmetry breaking and the Josephson
  current in a ballistic superconductor–quantum wire–superconductor
  junction}},\ }\href {https://doi.org/10.1063/1.1739160} {\bibfield  {journal}
  {\bibinfo  {journal} {Low Temp. Phys.}\ }\textbf {\bibinfo {volume} {30}},\
  \bibinfo {pages} {398} (\bibinfo {year} {2004})}\BibitemShut {NoStop}%
\bibitem [{\citenamefont {Buzdin}(2008)}]{Buzdin2008}%
  \BibitemOpen
  \bibfield  {author} {\bibinfo {author} {\bibfnamefont {A.}~\bibnamefont
  {Buzdin}},\ }\bibfield  {title} {\bibinfo {title} {{Direct Coupling Between
  Magnetism and Superconducting Current in the Josephson
  ${\ensuremath{\varphi}}_{0}$ Junction}},\ }\href
  {https://doi.org/10.1103/PhysRevLett.101.107005} {\bibfield  {journal}
  {\bibinfo  {journal} {Phys. Rev. Lett.}\ }\textbf {\bibinfo {volume} {101}},\
  \bibinfo {pages} {107005} (\bibinfo {year} {2008})}\BibitemShut {NoStop}%
\bibitem [{\citenamefont {Reynoso}\ \emph {et~al.}(2008)\citenamefont
  {Reynoso}, \citenamefont {Usaj}, \citenamefont {Balseiro}, \citenamefont
  {Feinberg},\ and\ \citenamefont {Avignon}}]{Reynoso2008}%
  \BibitemOpen
  \bibfield  {author} {\bibinfo {author} {\bibfnamefont {A.~A.}\ \bibnamefont
  {Reynoso}}, \bibinfo {author} {\bibfnamefont {G.}~\bibnamefont {Usaj}},
  \bibinfo {author} {\bibfnamefont {C.~A.}\ \bibnamefont {Balseiro}}, \bibinfo
  {author} {\bibfnamefont {D.}~\bibnamefont {Feinberg}},\ and\ \bibinfo
  {author} {\bibfnamefont {M.}~\bibnamefont {Avignon}},\ }\bibfield  {title}
  {\bibinfo {title} {{Anomalous Josephson Current in Junctions with Spin
  Polarizing Quantum Point Contacts}},\ }\href
  {https://doi.org/10.1103/PhysRevLett.101.107001} {\bibfield  {journal}
  {\bibinfo  {journal} {Phys. Rev. Lett.}\ }\textbf {\bibinfo {volume} {101}},\
  \bibinfo {pages} {107001} (\bibinfo {year} {2008})}\BibitemShut {NoStop}%
\bibitem [{\citenamefont {Zazunov}\ \emph {et~al.}(2009)\citenamefont
  {Zazunov}, \citenamefont {Egger}, \citenamefont {Jonckheere},\ and\
  \citenamefont {Martin}}]{Zazunov2009}%
  \BibitemOpen
  \bibfield  {author} {\bibinfo {author} {\bibfnamefont {A.}~\bibnamefont
  {Zazunov}}, \bibinfo {author} {\bibfnamefont {R.}~\bibnamefont {Egger}},
  \bibinfo {author} {\bibfnamefont {T.}~\bibnamefont {Jonckheere}},\ and\
  \bibinfo {author} {\bibfnamefont {T.}~\bibnamefont {Martin}},\ }\bibfield
  {title} {\bibinfo {title} {{Anomalous Josephson Current through a Spin-Orbit
  Coupled Quantum Dot}},\ }\href
  {https://doi.org/10.1103/PhysRevLett.103.147004} {\bibfield  {journal}
  {\bibinfo  {journal} {Phys. Rev. Lett.}\ }\textbf {\bibinfo {volume} {103}},\
  \bibinfo {pages} {147004} (\bibinfo {year} {2009})}\BibitemShut {NoStop}%
\bibitem [{\citenamefont {Liu}\ and\ \citenamefont
  {Chan}(2010{\natexlab{a}})}]{Liu2010a}%
  \BibitemOpen
  \bibfield  {author} {\bibinfo {author} {\bibfnamefont {J.~F.}\ \bibnamefont
  {Liu}}\ and\ \bibinfo {author} {\bibfnamefont {K.~S.}\ \bibnamefont {Chan}},\
  }\bibfield  {title} {\bibinfo {title} {{Relation between symmetry breaking
  and the anomalous Josephson effect}},\ }\href
  {https://doi.org/10.1103/PhysRevB.82.125305} {\bibfield  {journal} {\bibinfo
  {journal} {Phys. Rev. B}\ }\textbf {\bibinfo {volume} {82}},\ \bibinfo
  {pages} {1} (\bibinfo {year} {2010}{\natexlab{a}})}\BibitemShut {NoStop}%
\bibitem [{\citenamefont {Liu}\ and\ \citenamefont
  {Chan}(2010{\natexlab{b}})}]{Liu2010}%
  \BibitemOpen
  \bibfield  {author} {\bibinfo {author} {\bibfnamefont {J.-F.}\ \bibnamefont
  {Liu}}\ and\ \bibinfo {author} {\bibfnamefont {K.~S.}\ \bibnamefont {Chan}},\
  }\bibfield  {title} {\bibinfo {title} {{Anomalous Josephson current through a
  ferromagnetic trilayer junction}},\ }\href
  {https://doi.org/10.1103/PhysRevB.82.184533} {\bibfield  {journal} {\bibinfo
  {journal} {Phys. Rev. B}\ }\textbf {\bibinfo {volume} {82}},\ \bibinfo
  {pages} {184533} (\bibinfo {year} {2010}{\natexlab{b}})}\BibitemShut
  {NoStop}%
\bibitem [{\citenamefont {Liu}\ \emph {et~al.}(2011)\citenamefont {Liu},
  \citenamefont {Sum~Chan},\ and\ \citenamefont {Wang}}]{Liu2011}%
  \BibitemOpen
  \bibfield  {author} {\bibinfo {author} {\bibfnamefont {J.-F.}\ \bibnamefont
  {Liu}}, \bibinfo {author} {\bibfnamefont {K.}~\bibnamefont {Sum~Chan}},\ and\
  \bibinfo {author} {\bibfnamefont {J.}~\bibnamefont {Wang}},\ }\bibfield
  {title} {\bibinfo {title} {Anomalous josephson current through a
  ferromagnet–semiconductor hybrid structure},\ }\href
  {https://doi.org/10.1143/JPSJ.80.124708} {\bibfield  {journal} {\bibinfo
  {journal} {Journal of the Physical Society of Japan}\ }\textbf {\bibinfo
  {volume} {80}},\ \bibinfo {pages} {124708} (\bibinfo {year} {2011})},\
  \Eprint {https://arxiv.org/abs/https://doi.org/10.1143/JPSJ.80.124708}
  {https://doi.org/10.1143/JPSJ.80.124708} \BibitemShut {NoStop}%
\bibitem [{\citenamefont {Reynoso}\ \emph {et~al.}(2012)\citenamefont
  {Reynoso}, \citenamefont {Usaj}, \citenamefont {Balseiro}, \citenamefont
  {Feinberg},\ and\ \citenamefont {Avignon}}]{Reynoso2012}%
  \BibitemOpen
  \bibfield  {author} {\bibinfo {author} {\bibfnamefont {A.~A.}\ \bibnamefont
  {Reynoso}}, \bibinfo {author} {\bibfnamefont {G.}~\bibnamefont {Usaj}},
  \bibinfo {author} {\bibfnamefont {C.~A.}\ \bibnamefont {Balseiro}}, \bibinfo
  {author} {\bibfnamefont {D.}~\bibnamefont {Feinberg}},\ and\ \bibinfo
  {author} {\bibfnamefont {M.}~\bibnamefont {Avignon}},\ }\bibfield  {title}
  {\bibinfo {title} {{Spin-orbit-induced chirality of Andreev states in
  Josephson junctions}},\ }\href {https://doi.org/10.1103/PhysRevB.86.214519}
  {\bibfield  {journal} {\bibinfo  {journal} {Phys. Rev. B}\ }\textbf {\bibinfo
  {volume} {86}},\ \bibinfo {pages} {214519} (\bibinfo {year}
  {2012})}\BibitemShut {NoStop}%
\bibitem [{\citenamefont {Yokoyama}\ \emph {et~al.}(2013)\citenamefont
  {Yokoyama}, \citenamefont {Eto},\ and\ \citenamefont
  {V.~Nazarov}}]{YokoyamaJPSJ2013}%
  \BibitemOpen
  \bibfield  {author} {\bibinfo {author} {\bibfnamefont {T.}~\bibnamefont
  {Yokoyama}}, \bibinfo {author} {\bibfnamefont {M.}~\bibnamefont {Eto}},\ and\
  \bibinfo {author} {\bibfnamefont {Y.}~\bibnamefont {V.~Nazarov}},\ }\bibfield
   {title} {\bibinfo {title} {{Josephson Current through Semiconductor Nanowire
  with Spin–Orbit Interaction in Magnetic Field}},\ }\href
  {https://doi.org/10.7566/JPSJ.82.054703} {\bibfield  {journal} {\bibinfo
  {journal} {Journal of the Physical Society of Japan}\ }\textbf {\bibinfo
  {volume} {82}},\ \bibinfo {pages} {054703} (\bibinfo {year} {2013})},\
  \Eprint {https://arxiv.org/abs/https://doi.org/10.7566/JPSJ.82.054703}
  {https://doi.org/10.7566/JPSJ.82.054703} \BibitemShut {NoStop}%
\bibitem [{\citenamefont {Brunetti}\ \emph {et~al.}(2013)\citenamefont
  {Brunetti}, \citenamefont {Zazunov}, \citenamefont {Kundu},\ and\
  \citenamefont {Egger}}]{Brunetti2013}%
  \BibitemOpen
  \bibfield  {author} {\bibinfo {author} {\bibfnamefont {A.}~\bibnamefont
  {Brunetti}}, \bibinfo {author} {\bibfnamefont {A.}~\bibnamefont {Zazunov}},
  \bibinfo {author} {\bibfnamefont {A.}~\bibnamefont {Kundu}},\ and\ \bibinfo
  {author} {\bibfnamefont {R.}~\bibnamefont {Egger}},\ }\bibfield  {title}
  {\bibinfo {title} {{Anomalous Josephson current, incipient time-reversal
  symmetry breaking, and Majorana bound states in interacting multilevel
  dots}},\ }\href {https://doi.org/10.1103/PhysRevB.88.144515} {\bibfield
  {journal} {\bibinfo  {journal} {Phys. Rev. B}\ }\textbf {\bibinfo {volume}
  {88}},\ \bibinfo {pages} {144515} (\bibinfo {year} {2013})}\BibitemShut
  {NoStop}%
\bibitem [{\citenamefont {Yokoyama}\ \emph {et~al.}(2014)\citenamefont
  {Yokoyama}, \citenamefont {Eto},\ and\ \citenamefont
  {Nazarov}}]{Yokoyama2014}%
  \BibitemOpen
  \bibfield  {author} {\bibinfo {author} {\bibfnamefont {T.}~\bibnamefont
  {Yokoyama}}, \bibinfo {author} {\bibfnamefont {M.}~\bibnamefont {Eto}},\ and\
  \bibinfo {author} {\bibfnamefont {Y.~V.}\ \bibnamefont {Nazarov}},\
  }\bibfield  {title} {\bibinfo {title} {{Anomalous Josephson effect induced by
  spin-orbit interaction and Zeeman effect in semiconductor nanowires}},\
  }\href {https://doi.org/10.1103/PhysRevB.89.195407} {\bibfield  {journal}
  {\bibinfo  {journal} {Phys. Rev. B}\ }\textbf {\bibinfo {volume} {89}},\
  \bibinfo {pages} {195407} (\bibinfo {year} {2014})}\BibitemShut {NoStop}%
\bibitem [{\citenamefont {Shen}\ \emph {et~al.}(2014)\citenamefont {Shen},
  \citenamefont {Vignale},\ and\ \citenamefont {Raimondi}}]{Shen2014}%
  \BibitemOpen
  \bibfield  {author} {\bibinfo {author} {\bibfnamefont {K.}~\bibnamefont
  {Shen}}, \bibinfo {author} {\bibfnamefont {G.}~\bibnamefont {Vignale}},\ and\
  \bibinfo {author} {\bibfnamefont {R.}~\bibnamefont {Raimondi}},\ }\bibfield
  {title} {\bibinfo {title} {{Microscopic Theory of the Inverse Edelstein
  Effect}},\ }\href {https://doi.org/10.1103/PhysRevLett.112.096601} {\bibfield
   {journal} {\bibinfo  {journal} {Phys. Rev. Lett.}\ }\textbf {\bibinfo
  {volume} {112}},\ \bibinfo {pages} {096601} (\bibinfo {year}
  {2014})}\BibitemShut {NoStop}%
\bibitem [{\citenamefont {Konschelle}\ \emph {et~al.}(2015)\citenamefont
  {Konschelle}, \citenamefont {Tokatly},\ and\ \citenamefont
  {Bergeret}}]{Konschelle2015}%
  \BibitemOpen
  \bibfield  {author} {\bibinfo {author} {\bibfnamefont {F.}~\bibnamefont
  {Konschelle}}, \bibinfo {author} {\bibfnamefont {I.~V.}\ \bibnamefont
  {Tokatly}},\ and\ \bibinfo {author} {\bibfnamefont {F.~S.}\ \bibnamefont
  {Bergeret}},\ }\bibfield  {title} {\bibinfo {title} {{Theory of the
  spin-galvanic effect and the anomalous phase shift
  ${\ensuremath{\varphi}}_{0}$ in superconductors and Josephson junctions with
  intrinsic spin-orbit coupling}},\ }\href
  {https://doi.org/10.1103/PhysRevB.92.125443} {\bibfield  {journal} {\bibinfo
  {journal} {Phys. Rev. B}\ }\textbf {\bibinfo {volume} {92}},\ \bibinfo
  {pages} {125443} (\bibinfo {year} {2015})}\BibitemShut {NoStop}%
\bibitem [{\citenamefont {Szombati}\ \emph {et~al.}(2016)\citenamefont
  {Szombati}, \citenamefont {Nadj-Perge}, \citenamefont {Car}, \citenamefont
  {Plissard}, \citenamefont {Bakkers},\ and\ \citenamefont
  {Kouwenhoven}}]{Szombati2016}%
  \BibitemOpen
  \bibfield  {author} {\bibinfo {author} {\bibfnamefont {D.~B.}\ \bibnamefont
  {Szombati}}, \bibinfo {author} {\bibfnamefont {S.}~\bibnamefont
  {Nadj-Perge}}, \bibinfo {author} {\bibfnamefont {D.}~\bibnamefont {Car}},
  \bibinfo {author} {\bibfnamefont {S.~R.}\ \bibnamefont {Plissard}}, \bibinfo
  {author} {\bibfnamefont {E.~P. A.~M.}\ \bibnamefont {Bakkers}},\ and\
  \bibinfo {author} {\bibfnamefont {L.~P.}\ \bibnamefont {Kouwenhoven}},\
  }\bibfield  {title} {\bibinfo {title} {{Josephson $\varphi_0$-junction in
  nanowire quantum dots}},\ }\href {https://doi.org/10.1038/nphys3742}
  {\bibfield  {journal} {\bibinfo  {journal} {Nature Physics}\ }\textbf
  {\bibinfo {volume} {12}},\ \bibinfo {pages} {568} (\bibinfo {year}
  {2016})}\BibitemShut {NoStop}%
\bibitem [{\citenamefont {Assouline}\ \emph {et~al.}(2019)\citenamefont
  {Assouline}, \citenamefont {Feuillet-Palma}, \citenamefont {Bergeal},
  \citenamefont {Zhang}, \citenamefont {Mottaghizadeh}, \citenamefont
  {Zimmers}, \citenamefont {Lhuillier}, \citenamefont {Eddrie}, \citenamefont
  {Atkinson}, \citenamefont {Aprili},\ and\ \citenamefont
  {Aubin}}]{Assouline2019}%
  \BibitemOpen
  \bibfield  {author} {\bibinfo {author} {\bibfnamefont {A.}~\bibnamefont
  {Assouline}}, \bibinfo {author} {\bibfnamefont {C.}~\bibnamefont
  {Feuillet-Palma}}, \bibinfo {author} {\bibfnamefont {N.}~\bibnamefont
  {Bergeal}}, \bibinfo {author} {\bibfnamefont {T.}~\bibnamefont {Zhang}},
  \bibinfo {author} {\bibfnamefont {A.}~\bibnamefont {Mottaghizadeh}}, \bibinfo
  {author} {\bibfnamefont {A.}~\bibnamefont {Zimmers}}, \bibinfo {author}
  {\bibfnamefont {E.}~\bibnamefont {Lhuillier}}, \bibinfo {author}
  {\bibfnamefont {M.}~\bibnamefont {Eddrie}}, \bibinfo {author} {\bibfnamefont
  {P.}~\bibnamefont {Atkinson}}, \bibinfo {author} {\bibfnamefont
  {M.}~\bibnamefont {Aprili}},\ and\ \bibinfo {author} {\bibfnamefont
  {H.}~\bibnamefont {Aubin}},\ }\bibfield  {title} {\bibinfo {title}
  {{Spin-Orbit induced phase-shift in Bi2Se3 Josephson junctions}},\ }\href
  {https://doi.org/10.1038/s41467-018-08022-y} {\bibfield  {journal} {\bibinfo
  {journal} {Nature Communications}\ }\textbf {\bibinfo {volume} {10}},\
  \bibinfo {pages} {126} (\bibinfo {year} {2019})}\BibitemShut {NoStop}%
\bibitem [{\citenamefont {Mayer}\ \emph {et~al.}(2020)\citenamefont {Mayer},
  \citenamefont {Dartiailh}, \citenamefont {Yuan}, \citenamefont
  {Wickramasinghe}, \citenamefont {Rossi},\ and\ \citenamefont
  {Shabani}}]{Mayer2020b}%
  \BibitemOpen
  \bibfield  {author} {\bibinfo {author} {\bibfnamefont {W.}~\bibnamefont
  {Mayer}}, \bibinfo {author} {\bibfnamefont {M.~C.}\ \bibnamefont
  {Dartiailh}}, \bibinfo {author} {\bibfnamefont {J.}~\bibnamefont {Yuan}},
  \bibinfo {author} {\bibfnamefont {K.~S.}\ \bibnamefont {Wickramasinghe}},
  \bibinfo {author} {\bibfnamefont {E.}~\bibnamefont {Rossi}},\ and\ \bibinfo
  {author} {\bibfnamefont {J.}~\bibnamefont {Shabani}},\ }\bibfield  {title}
  {\bibinfo {title} {{Gate controlled anomalous phase shift in Al/InAs
  Josephson junctions}},\ }\href {https://doi.org/10.1038/s41467-019-14094-1}
  {\bibfield  {journal} {\bibinfo  {journal} {Nature Communications}\ }\textbf
  {\bibinfo {volume} {11}},\ \bibinfo {pages} {212} (\bibinfo {year}
  {2020})}\BibitemShut {NoStop}%
\bibitem [{\citenamefont {Strambini}\ \emph {et~al.}(2020)\citenamefont
  {Strambini}, \citenamefont {Iorio}, \citenamefont {Durante}, \citenamefont
  {Citro}, \citenamefont {Sanz-Fern{\'a}ndez}, \citenamefont {Guarcello},
  \citenamefont {Tokatly}, \citenamefont {Braggio}, \citenamefont {Rocci},
  \citenamefont {Ligato}, \citenamefont {Zannier}, \citenamefont {Sorba},
  \citenamefont {Bergeret},\ and\ \citenamefont {Giazotto}}]{Strambini2020}%
  \BibitemOpen
  \bibfield  {author} {\bibinfo {author} {\bibfnamefont {E.}~\bibnamefont
  {Strambini}}, \bibinfo {author} {\bibfnamefont {A.}~\bibnamefont {Iorio}},
  \bibinfo {author} {\bibfnamefont {O.}~\bibnamefont {Durante}}, \bibinfo
  {author} {\bibfnamefont {R.}~\bibnamefont {Citro}}, \bibinfo {author}
  {\bibfnamefont {C.}~\bibnamefont {Sanz-Fern{\'a}ndez}}, \bibinfo {author}
  {\bibfnamefont {C.}~\bibnamefont {Guarcello}}, \bibinfo {author}
  {\bibfnamefont {I.~V.}\ \bibnamefont {Tokatly}}, \bibinfo {author}
  {\bibfnamefont {A.}~\bibnamefont {Braggio}}, \bibinfo {author} {\bibfnamefont
  {M.}~\bibnamefont {Rocci}}, \bibinfo {author} {\bibfnamefont
  {N.}~\bibnamefont {Ligato}}, \bibinfo {author} {\bibfnamefont
  {V.}~\bibnamefont {Zannier}}, \bibinfo {author} {\bibfnamefont
  {L.}~\bibnamefont {Sorba}}, \bibinfo {author} {\bibfnamefont {F.~S.}\
  \bibnamefont {Bergeret}},\ and\ \bibinfo {author} {\bibfnamefont
  {F.}~\bibnamefont {Giazotto}},\ }\bibfield  {title} {\bibinfo {title} {{A
  Josephson phase battery}},\ }\href
  {https://doi.org/10.1038/s41565-020-0712-7} {\bibfield  {journal} {\bibinfo
  {journal} {Nature Nanotechnology}\ }\textbf {\bibinfo {volume} {15}},\
  \bibinfo {pages} {656} (\bibinfo {year} {2020})}\BibitemShut {NoStop}%
\bibitem [{\citenamefont {Baumgartner}\ \emph {et~al.}(2021)\citenamefont
  {Baumgartner}, \citenamefont {Fuchs}, \citenamefont {Fr\'esz}, \citenamefont
  {Reinhardt}, \citenamefont {Gronin}, \citenamefont {Gardner}, \citenamefont
  {Manfra}, \citenamefont {Paradiso},\ and\ \citenamefont
  {Strunk}}]{baumgartner2020}%
  \BibitemOpen
  \bibfield  {author} {\bibinfo {author} {\bibfnamefont {C.}~\bibnamefont
  {Baumgartner}}, \bibinfo {author} {\bibfnamefont {L.}~\bibnamefont {Fuchs}},
  \bibinfo {author} {\bibfnamefont {L.}~\bibnamefont {Fr\'esz}}, \bibinfo
  {author} {\bibfnamefont {S.}~\bibnamefont {Reinhardt}}, \bibinfo {author}
  {\bibfnamefont {S.}~\bibnamefont {Gronin}}, \bibinfo {author} {\bibfnamefont
  {G.~C.}\ \bibnamefont {Gardner}}, \bibinfo {author} {\bibfnamefont {M.~J.}\
  \bibnamefont {Manfra}}, \bibinfo {author} {\bibfnamefont {N.}~\bibnamefont
  {Paradiso}},\ and\ \bibinfo {author} {\bibfnamefont {C.}~\bibnamefont
  {Strunk}},\ }\bibfield  {title} {\bibinfo {title} {{Josephson Inductance as a
  Probe for Highly Ballistic Semiconductor-Superconductor Weak Links}},\ }\href
  {https://doi.org/10.1103/PhysRevLett.126.037001} {\bibfield  {journal}
  {\bibinfo  {journal} {Phys. Rev. Lett.}\ }\textbf {\bibinfo {volume} {126}},\
  \bibinfo {pages} {037001} (\bibinfo {year} {2021})}\BibitemShut {NoStop}%
\bibitem [{\citenamefont {De~Gennes}(1989)}]{DeGennes1989}%
  \BibitemOpen
  \bibfield  {author} {\bibinfo {author} {\bibfnamefont {P.~G.}\ \bibnamefont
  {De~Gennes}},\ }\href@noop {} {\emph {\bibinfo {title} {{Superconductivity of
  Metals and Alloys}}}}\ (\bibinfo  {publisher} {Addison Wesley, Redwood
  City},\ \bibinfo {year} {1989})\BibitemShut {NoStop}%
\bibitem [{\citenamefont {Li}\ \emph {et~al.}(2019)\citenamefont {Li},
  \citenamefont {de~Ronde}, \citenamefont {de~Boer}, \citenamefont {Ridderbos},
  \citenamefont {Zwanenburg}, \citenamefont {Huang}, \citenamefont {Golubov},\
  and\ \citenamefont {Brinkman}}]{LiPRL2019}%
  \BibitemOpen
  \bibfield  {author} {\bibinfo {author} {\bibfnamefont {C.}~\bibnamefont
  {Li}}, \bibinfo {author} {\bibfnamefont {B.}~\bibnamefont {de~Ronde}},
  \bibinfo {author} {\bibfnamefont {J.}~\bibnamefont {de~Boer}}, \bibinfo
  {author} {\bibfnamefont {J.}~\bibnamefont {Ridderbos}}, \bibinfo {author}
  {\bibfnamefont {F.}~\bibnamefont {Zwanenburg}}, \bibinfo {author}
  {\bibfnamefont {Y.}~\bibnamefont {Huang}}, \bibinfo {author} {\bibfnamefont
  {A.}~\bibnamefont {Golubov}},\ and\ \bibinfo {author} {\bibfnamefont
  {A.}~\bibnamefont {Brinkman}},\ }\bibfield  {title} {\bibinfo {title}
  {{Zeeman-Effect-Induced $0\text{\ensuremath{-}}\ensuremath{\pi}$ Transitions
  in Ballistic Dirac Semimetal Josephson Junctions}},\ }\href
  {https://doi.org/10.1103/PhysRevLett.123.026802} {\bibfield  {journal}
  {\bibinfo  {journal} {Phys. Rev. Lett.}\ }\textbf {\bibinfo {volume} {123}},\
  \bibinfo {pages} {026802} (\bibinfo {year} {2019})}\BibitemShut {NoStop}%
\bibitem [{\citenamefont {Hart}\ \emph {et~al.}(2017)\citenamefont {Hart},
  \citenamefont {Ren}, \citenamefont {Kosowsky}, \citenamefont {Ben-Shach},
  \citenamefont {Leubner}, \citenamefont {Br{\"u}ne}, \citenamefont {Buhmann},
  \citenamefont {Molenkamp}, \citenamefont {Halperin},\ and\ \citenamefont
  {Yacoby}}]{Hart2017}%
  \BibitemOpen
  \bibfield  {author} {\bibinfo {author} {\bibfnamefont {S.}~\bibnamefont
  {Hart}}, \bibinfo {author} {\bibfnamefont {H.}~\bibnamefont {Ren}}, \bibinfo
  {author} {\bibfnamefont {M.}~\bibnamefont {Kosowsky}}, \bibinfo {author}
  {\bibfnamefont {G.}~\bibnamefont {Ben-Shach}}, \bibinfo {author}
  {\bibfnamefont {P.}~\bibnamefont {Leubner}}, \bibinfo {author} {\bibfnamefont
  {C.}~\bibnamefont {Br{\"u}ne}}, \bibinfo {author} {\bibfnamefont
  {H.}~\bibnamefont {Buhmann}}, \bibinfo {author} {\bibfnamefont {L.~W.}\
  \bibnamefont {Molenkamp}}, \bibinfo {author} {\bibfnamefont {B.~I.}\
  \bibnamefont {Halperin}},\ and\ \bibinfo {author} {\bibfnamefont
  {A.}~\bibnamefont {Yacoby}},\ }\bibfield  {title} {\bibinfo {title}
  {{Controlled finite momentum pairing and spatially varying order parameter in
  proximitized HgTe quantum wells}},\ }\href
  {https://doi.org/10.1038/nphys3877} {\bibfield  {journal} {\bibinfo
  {journal} {Nature Physics}\ }\textbf {\bibinfo {volume} {13}},\ \bibinfo
  {pages} {87} (\bibinfo {year} {2017})}\BibitemShut {NoStop}%
\bibitem [{\citenamefont {Chen}\ \emph {et~al.}(2018)\citenamefont {Chen},
  \citenamefont {Park}, \citenamefont {Gill}, \citenamefont {Xiao},
  \citenamefont {Reig-i Plessis}, \citenamefont {MacDougall}, \citenamefont
  {Gilbert},\ and\ \citenamefont {Mason}}]{ChenNatComm2018}%
  \BibitemOpen
  \bibfield  {author} {\bibinfo {author} {\bibfnamefont {A.~Q.}\ \bibnamefont
  {Chen}}, \bibinfo {author} {\bibfnamefont {M.~J.}\ \bibnamefont {Park}},
  \bibinfo {author} {\bibfnamefont {S.~T.}\ \bibnamefont {Gill}}, \bibinfo
  {author} {\bibfnamefont {Y.}~\bibnamefont {Xiao}}, \bibinfo {author}
  {\bibfnamefont {D.}~\bibnamefont {Reig-i Plessis}}, \bibinfo {author}
  {\bibfnamefont {G.~J.}\ \bibnamefont {MacDougall}}, \bibinfo {author}
  {\bibfnamefont {M.~J.}\ \bibnamefont {Gilbert}},\ and\ \bibinfo {author}
  {\bibfnamefont {N.}~\bibnamefont {Mason}},\ }\bibfield  {title} {\bibinfo
  {title} {{Finite momentum Cooper pairing in three-dimensional topological
  insulator Josephson junctions}},\ }\href
  {https://doi.org/10.1038/s41467-018-05993-w} {\bibfield  {journal} {\bibinfo
  {journal} {Nature Communications}\ }\textbf {\bibinfo {volume} {9}},\
  \bibinfo {pages} {3478} (\bibinfo {year} {2018})}\BibitemShut {NoStop}%
\bibitem [{\citenamefont {Ke}\ \emph {et~al.}(2019)\citenamefont {Ke},
  \citenamefont {Moehle}, \citenamefont {de~Vries}, \citenamefont {Thomas},
  \citenamefont {Metti}, \citenamefont {Guinn}, \citenamefont {Kallaher},
  \citenamefont {Lodari}, \citenamefont {Scappucci}, \citenamefont {Wang},
  \citenamefont {Diaz}, \citenamefont {Gardner}, \citenamefont {Manfra},\ and\
  \citenamefont {Goswami}}]{Ke2019}%
  \BibitemOpen
  \bibfield  {author} {\bibinfo {author} {\bibfnamefont {C.~T.}\ \bibnamefont
  {Ke}}, \bibinfo {author} {\bibfnamefont {C.~M.}\ \bibnamefont {Moehle}},
  \bibinfo {author} {\bibfnamefont {F.~K.}\ \bibnamefont {de~Vries}}, \bibinfo
  {author} {\bibfnamefont {C.}~\bibnamefont {Thomas}}, \bibinfo {author}
  {\bibfnamefont {S.}~\bibnamefont {Metti}}, \bibinfo {author} {\bibfnamefont
  {C.~R.}\ \bibnamefont {Guinn}}, \bibinfo {author} {\bibfnamefont
  {R.}~\bibnamefont {Kallaher}}, \bibinfo {author} {\bibfnamefont
  {M.}~\bibnamefont {Lodari}}, \bibinfo {author} {\bibfnamefont
  {G.}~\bibnamefont {Scappucci}}, \bibinfo {author} {\bibfnamefont
  {T.}~\bibnamefont {Wang}}, \bibinfo {author} {\bibfnamefont {R.~E.}\
  \bibnamefont {Diaz}}, \bibinfo {author} {\bibfnamefont {G.~C.}\ \bibnamefont
  {Gardner}}, \bibinfo {author} {\bibfnamefont {M.~J.}\ \bibnamefont
  {Manfra}},\ and\ \bibinfo {author} {\bibfnamefont {S.}~\bibnamefont
  {Goswami}},\ }\bibfield  {title} {\bibinfo {title} {Ballistic
  superconductivity and tunable $\pi$--junctions in insb quantum wells},\
  }\href {https://doi.org/10.1038/s41467-019-11742-4} {\bibfield  {journal}
  {\bibinfo  {journal} {Nature Communications}\ }\textbf {\bibinfo {volume}
  {10}},\ \bibinfo {pages} {3764} (\bibinfo {year} {2019})}\BibitemShut
  {NoStop}%
\bibitem [{\citenamefont {Whiticar}\ \emph {et~al.}(2021)\citenamefont
  {Whiticar}, \citenamefont {Fornieri}, \citenamefont {Banerjee}, \citenamefont
  {Drachmann}, \citenamefont {Gronin}, \citenamefont {Gardner}, \citenamefont
  {Lindemann}, \citenamefont {Manfra},\ and\ \citenamefont
  {Marcus}}]{WhiticarPRB2021}%
  \BibitemOpen
  \bibfield  {author} {\bibinfo {author} {\bibfnamefont {A.~M.}\ \bibnamefont
  {Whiticar}}, \bibinfo {author} {\bibfnamefont {A.}~\bibnamefont {Fornieri}},
  \bibinfo {author} {\bibfnamefont {A.}~\bibnamefont {Banerjee}}, \bibinfo
  {author} {\bibfnamefont {A.~C.~C.}\ \bibnamefont {Drachmann}}, \bibinfo
  {author} {\bibfnamefont {S.}~\bibnamefont {Gronin}}, \bibinfo {author}
  {\bibfnamefont {G.~C.}\ \bibnamefont {Gardner}}, \bibinfo {author}
  {\bibfnamefont {T.}~\bibnamefont {Lindemann}}, \bibinfo {author}
  {\bibfnamefont {M.~J.}\ \bibnamefont {Manfra}},\ and\ \bibinfo {author}
  {\bibfnamefont {C.~M.}\ \bibnamefont {Marcus}},\ }\bibfield  {title}
  {\bibinfo {title} {Zeeman-driven parity transitions in an andreev quantum
  dot},\ }\href {https://doi.org/10.1103/PhysRevB.103.245308} {\bibfield
  {journal} {\bibinfo  {journal} {Phys. Rev. B}\ }\textbf {\bibinfo {volume}
  {103}},\ \bibinfo {pages} {245308} (\bibinfo {year} {2021})}\BibitemShut
  {NoStop}%
\bibitem [{\citenamefont {Shin}\ \emph {et~al.}(2021)\citenamefont {Shin},
  \citenamefont {Son}, \citenamefont {Yun}, \citenamefont {Park}, \citenamefont
  {Zhang}, \citenamefont {Shin}, \citenamefont {Park},\ and\ \citenamefont
  {Kim}}]{shin2021}%
  \BibitemOpen
  \bibfield  {author} {\bibinfo {author} {\bibfnamefont {J.}~\bibnamefont
  {Shin}}, \bibinfo {author} {\bibfnamefont {S.}~\bibnamefont {Son}}, \bibinfo
  {author} {\bibfnamefont {J.}~\bibnamefont {Yun}}, \bibinfo {author}
  {\bibfnamefont {G.}~\bibnamefont {Park}}, \bibinfo {author} {\bibfnamefont
  {K.}~\bibnamefont {Zhang}}, \bibinfo {author} {\bibfnamefont {Y.~J.}\
  \bibnamefont {Shin}}, \bibinfo {author} {\bibfnamefont {J.-G.}\ \bibnamefont
  {Park}},\ and\ \bibinfo {author} {\bibfnamefont {D.}~\bibnamefont {Kim}},\
  }\href@noop {} {\bibinfo {title} {Magnetic proximity-induced superconducting
  diode effect and infinite magnetoresistance in van der waals
  heterostructure}} (\bibinfo {year} {2021}),\ \Eprint
  {https://arxiv.org/abs/2111.05627} {arXiv:2111.05627 [cond-mat.supr-con]}
  \BibitemShut {NoStop}%
\bibitem [{\citenamefont {Suri}\ \emph {et~al.}(2022)\citenamefont {Suri},
  \citenamefont {Kamra}, \citenamefont {Meier}, \citenamefont {Kronseder},
  \citenamefont {Belzig}, \citenamefont {Back},\ and\ \citenamefont
  {Strunk}}]{Suri2022}%
  \BibitemOpen
  \bibfield  {author} {\bibinfo {author} {\bibfnamefont {D.}~\bibnamefont
  {Suri}}, \bibinfo {author} {\bibfnamefont {A.}~\bibnamefont {Kamra}},
  \bibinfo {author} {\bibfnamefont {T.~N.~G.}\ \bibnamefont {Meier}}, \bibinfo
  {author} {\bibfnamefont {M.}~\bibnamefont {Kronseder}}, \bibinfo {author}
  {\bibfnamefont {W.}~\bibnamefont {Belzig}}, \bibinfo {author} {\bibfnamefont
  {C.~H.}\ \bibnamefont {Back}},\ and\ \bibinfo {author} {\bibfnamefont
  {C.}~\bibnamefont {Strunk}},\ }\bibfield  {title} {\bibinfo {title}
  {Non-reciprocity of vortex-limited critical current in conventional
  superconducting micro-bridges},\ }\href {https://doi.org/10.1063/5.0109753}
  {\bibfield  {journal} {\bibinfo  {journal} {Applied Physics Letters}\
  }\textbf {\bibinfo {volume} {121}},\ \bibinfo {pages} {102601} (\bibinfo
  {year} {2022})},\ \Eprint
  {https://arxiv.org/abs/https://doi.org/10.1063/5.0109753}
  {https://doi.org/10.1063/5.0109753} \BibitemShut {NoStop}%
\bibitem [{\citenamefont {Hou}\ \emph {et~al.}(2022)\citenamefont {Hou},
  \citenamefont {Nichele}, \citenamefont {Chi}, \citenamefont {Lodesani},
  \citenamefont {Wu}, \citenamefont {Ritter}, \citenamefont {Haxell},
  \citenamefont {Davydova}, \citenamefont {Ilić}, \citenamefont {Bergeret},
  \citenamefont {Kamra}, \citenamefont {Fu}, \citenamefont {Lee},\ and\
  \citenamefont {Moodera}}]{Hou2022}%
  \BibitemOpen
  \bibfield  {author} {\bibinfo {author} {\bibfnamefont {Y.}~\bibnamefont
  {Hou}}, \bibinfo {author} {\bibfnamefont {F.}~\bibnamefont {Nichele}},
  \bibinfo {author} {\bibfnamefont {H.}~\bibnamefont {Chi}}, \bibinfo {author}
  {\bibfnamefont {A.}~\bibnamefont {Lodesani}}, \bibinfo {author}
  {\bibfnamefont {Y.}~\bibnamefont {Wu}}, \bibinfo {author} {\bibfnamefont
  {M.~F.}\ \bibnamefont {Ritter}}, \bibinfo {author} {\bibfnamefont {D.~Z.}\
  \bibnamefont {Haxell}}, \bibinfo {author} {\bibfnamefont {M.}~\bibnamefont
  {Davydova}}, \bibinfo {author} {\bibfnamefont {S.}~\bibnamefont {Ilić}},
  \bibinfo {author} {\bibfnamefont {F.~S.}\ \bibnamefont {Bergeret}}, \bibinfo
  {author} {\bibfnamefont {A.}~\bibnamefont {Kamra}}, \bibinfo {author}
  {\bibfnamefont {L.}~\bibnamefont {Fu}}, \bibinfo {author} {\bibfnamefont
  {P.~A.}\ \bibnamefont {Lee}},\ and\ \bibinfo {author} {\bibfnamefont {J.~S.}\
  \bibnamefont {Moodera}},\ }\href@noop {} {\bibinfo {title} {Ubiquitous
  superconducting diode effect in superconductor thin films}} (\bibinfo {year}
  {2022}),\ \Eprint {https://arxiv.org/abs/2205.09276} {arXiv:2205.09276
  [cond-mat.supr-con]} \BibitemShut {NoStop}%
\bibitem [{\citenamefont {Sundaresh}\ \emph {et~al.}(2022)\citenamefont
  {Sundaresh}, \citenamefont {Vayrynen}, \citenamefont {Lyanda-Geller},\ and\
  \citenamefont {Rokhinson}}]{Sundaresh2022}%
  \BibitemOpen
  \bibfield  {author} {\bibinfo {author} {\bibfnamefont {A.}~\bibnamefont
  {Sundaresh}}, \bibinfo {author} {\bibfnamefont {J.~I.}\ \bibnamefont
  {Vayrynen}}, \bibinfo {author} {\bibfnamefont {Y.}~\bibnamefont
  {Lyanda-Geller}},\ and\ \bibinfo {author} {\bibfnamefont {L.~P.}\
  \bibnamefont {Rokhinson}},\ }\href@noop {} {\bibinfo {title} {{Supercurrent
  non-reciprocity and vortex formation in superconductor heterostructures}}}
  (\bibinfo {year} {2022}),\ \Eprint {https://arxiv.org/abs/2207.03633}
  {arXiv:2207.03633 [cond-mat.supr-con]} \BibitemShut {NoStop}%
\bibitem [{\citenamefont {{\v{Z}}uti{\'{c}}}\ and\ \citenamefont
  {Valls}(2000)}]{Zutic2000}%
  \BibitemOpen
  \bibfield  {author} {\bibinfo {author} {\bibfnamefont {I.}~\bibnamefont
  {{\v{Z}}uti{\'{c}}}}\ and\ \bibinfo {author} {\bibfnamefont {O.~T.}\
  \bibnamefont {Valls}},\ }\bibfield  {title} {\bibinfo {title} {{Tunneling
  spectroscopy for ferromagnet/superconductor junctions}},\ }\href
  {http://link.aps.org/doi/10.1103/PhysRevB.61.1555} {\bibfield  {journal}
  {\bibinfo  {journal} {Phys. Rev. B}\ }\textbf {\bibinfo {volume} {61}},\
  \bibinfo {pages} {1555} (\bibinfo {year} {2000})}\BibitemShut {NoStop}%
\bibitem [{\citenamefont {Blonder}\ \emph {et~al.}(1982)\citenamefont
  {Blonder}, \citenamefont {Tinkham},\ and\ \citenamefont
  {Klapwijk}}]{Blonder1982}%
  \BibitemOpen
  \bibfield  {author} {\bibinfo {author} {\bibfnamefont {G.~E.}\ \bibnamefont
  {Blonder}}, \bibinfo {author} {\bibfnamefont {M.}~\bibnamefont {Tinkham}},\
  and\ \bibinfo {author} {\bibfnamefont {T.~M.}\ \bibnamefont {Klapwijk}},\
  }\bibfield  {title} {\bibinfo {title} {{Transition from metallic to tunneling
  regimes in superconducting microconstrictions: Excess current, charge
  imbalance, and supercurrent conversion}},\ }\href
  {https://doi.org/10.1103/PhysRevB.25.4515} {\bibfield  {journal} {\bibinfo
  {journal} {Phys. Rev. B}\ }\textbf {\bibinfo {volume} {25}},\ \bibinfo
  {pages} {4515} (\bibinfo {year} {1982})}\BibitemShut {NoStop}%
\bibitem [{\citenamefont {Dartiailh}\ \emph {et~al.}(2021)\citenamefont
  {Dartiailh}, \citenamefont {Mayer}, \citenamefont {Yuan}, \citenamefont
  {Wickramasinghe}, \citenamefont {Matos-Abiague}, \citenamefont
  {{\v{Z}}uti{\'{c}}},\ and\ \citenamefont {Shabani}}]{Dartiailh2021}%
  \BibitemOpen
  \bibfield  {author} {\bibinfo {author} {\bibfnamefont {M.~C.}\ \bibnamefont
  {Dartiailh}}, \bibinfo {author} {\bibfnamefont {W.}~\bibnamefont {Mayer}},
  \bibinfo {author} {\bibfnamefont {J.}~\bibnamefont {Yuan}}, \bibinfo {author}
  {\bibfnamefont {K.~S.}\ \bibnamefont {Wickramasinghe}}, \bibinfo {author}
  {\bibfnamefont {A.}~\bibnamefont {Matos-Abiague}}, \bibinfo {author}
  {\bibfnamefont {I.}~\bibnamefont {{\v{Z}}uti{\'{c}}}},\ and\ \bibinfo
  {author} {\bibfnamefont {J.}~\bibnamefont {Shabani}},\ }\bibfield  {title}
  {\bibinfo {title} {{Phase Signature of Topological Transition in Josephson
  Junctions}},\ }\href {https://doi.org/10.1103/PhysRevLett.126.036802}
  {\bibfield  {journal} {\bibinfo  {journal} {Phys. Rev. Lett.}\ }\textbf
  {\bibinfo {volume} {126}},\ \bibinfo {pages} {036802} (\bibinfo {year}
  {2021})}\BibitemShut {NoStop}%
\bibitem [{\citenamefont {Beenakker}\ and\ \citenamefont {van
  Houten}(1991)}]{BeenakkerPRL91}%
  \BibitemOpen
  \bibfield  {author} {\bibinfo {author} {\bibfnamefont {C.~W.~J.}\
  \bibnamefont {Beenakker}}\ and\ \bibinfo {author} {\bibfnamefont
  {H.}~\bibnamefont {van Houten}},\ }\bibfield  {title} {\bibinfo {title}
  {{Josephson current through a superconducting quantum point contact shorter
  than the coherence length}},\ }\href
  {https://doi.org/10.1103/PhysRevLett.66.3056} {\bibfield  {journal} {\bibinfo
   {journal} {Phys. Rev. Lett.}\ }\textbf {\bibinfo {volume} {66}},\ \bibinfo
  {pages} {3056} (\bibinfo {year} {1991})}\BibitemShut {NoStop}%
\bibitem [{\citenamefont {Della~Rocca}\ \emph {et~al.}(2007)\citenamefont
  {Della~Rocca}, \citenamefont {Chauvin}, \citenamefont {Huard}, \citenamefont
  {Pothier}, \citenamefont {Esteve},\ and\ \citenamefont
  {Urbina}}]{DellaRocca2007}%
  \BibitemOpen
  \bibfield  {author} {\bibinfo {author} {\bibfnamefont {M.~L.}\ \bibnamefont
  {Della~Rocca}}, \bibinfo {author} {\bibfnamefont {M.}~\bibnamefont
  {Chauvin}}, \bibinfo {author} {\bibfnamefont {B.}~\bibnamefont {Huard}},
  \bibinfo {author} {\bibfnamefont {H.}~\bibnamefont {Pothier}}, \bibinfo
  {author} {\bibfnamefont {D.}~\bibnamefont {Esteve}},\ and\ \bibinfo {author}
  {\bibfnamefont {C.}~\bibnamefont {Urbina}},\ }\bibfield  {title} {\bibinfo
  {title} {Measurement of the current-phase relation of superconducting atomic
  contacts},\ }\href {https://doi.org/10.1103/PhysRevLett.99.127005} {\bibfield
   {journal} {\bibinfo  {journal} {Phys. Rev. Lett.}\ }\textbf {\bibinfo
  {volume} {99}},\ \bibinfo {pages} {127005} (\bibinfo {year}
  {2007})}\BibitemShut {NoStop}%
\bibitem [{\citenamefont {Golubov}\ \emph {et~al.}(2004)\citenamefont
  {Golubov}, \citenamefont {Kupriyanov},\ and\ \citenamefont
  {Il'ichev}}]{RMPGolubov}%
  \BibitemOpen
  \bibfield  {author} {\bibinfo {author} {\bibfnamefont {A.~A.}\ \bibnamefont
  {Golubov}}, \bibinfo {author} {\bibfnamefont {M.~Y.}\ \bibnamefont
  {Kupriyanov}},\ and\ \bibinfo {author} {\bibfnamefont {E.}~\bibnamefont
  {Il'ichev}},\ }\bibfield  {title} {\bibinfo {title} {{The current-phase
  relation in Josephson junctions}},\ }\href
  {https://doi.org/10.1103/RevModPhys.76.411} {\bibfield  {journal} {\bibinfo
  {journal} {Rev. Mod. Phys.}\ }\textbf {\bibinfo {volume} {76}},\ \bibinfo
  {pages} {411} (\bibinfo {year} {2004})}\BibitemShut {NoStop}%
\bibitem [{Note1()}]{Note1}%
  \BibitemOpen
  \bibinfo {note} {Alternatively, one can use the temperature dependence of the
  zero-bias inductance to extract $\Delta ^{\ast }$, if the number of channels
  $N$ is not known~\cite {baumgartner2020}.}\BibitemShut {Stop}%
\bibitem [{\citenamefont {Pientka}\ \emph {et~al.}(2017)\citenamefont
  {Pientka}, \citenamefont {Keselman}, \citenamefont {Berg}, \citenamefont
  {Yacoby}, \citenamefont {Stern},\ and\ \citenamefont
  {Halperin}}]{Pientka2017}%
  \BibitemOpen
  \bibfield  {author} {\bibinfo {author} {\bibfnamefont {F.}~\bibnamefont
  {Pientka}}, \bibinfo {author} {\bibfnamefont {A.}~\bibnamefont {Keselman}},
  \bibinfo {author} {\bibfnamefont {E.}~\bibnamefont {Berg}}, \bibinfo {author}
  {\bibfnamefont {A.}~\bibnamefont {Yacoby}}, \bibinfo {author} {\bibfnamefont
  {A.}~\bibnamefont {Stern}},\ and\ \bibinfo {author} {\bibfnamefont {B.~I.}\
  \bibnamefont {Halperin}},\ }\bibfield  {title} {\bibinfo {title} {Topological
  superconductivity in a planar josephson junction},\ }\href
  {https://doi.org/10.1103/PhysRevX.7.021032} {\bibfield  {journal} {\bibinfo
  {journal} {Phys. Rev. X}\ }\textbf {\bibinfo {volume} {7}},\ \bibinfo {pages}
  {021032} (\bibinfo {year} {2017})}\BibitemShut {NoStop}%
\bibitem [{\citenamefont {Costa}\ \emph {et~al.}(2018)\citenamefont {Costa},
  \citenamefont {Fabian},\ and\ \citenamefont {Kochan}}]{Costa2018}%
  \BibitemOpen
  \bibfield  {author} {\bibinfo {author} {\bibfnamefont {A.}~\bibnamefont
  {Costa}}, \bibinfo {author} {\bibfnamefont {J.}~\bibnamefont {Fabian}},\ and\
  \bibinfo {author} {\bibfnamefont {D.}~\bibnamefont {Kochan}},\ }\bibfield
  {title} {\bibinfo {title} {Connection between zero-energy yu-shiba-rusinov
  states and $0\text{\ensuremath{-}}\ensuremath{\pi}$ transitions in magnetic
  josephson junctions},\ }\href {https://doi.org/10.1103/PhysRevB.98.134511}
  {\bibfield  {journal} {\bibinfo  {journal} {Phys. Rev. B}\ }\textbf {\bibinfo
  {volume} {98}},\ \bibinfo {pages} {134511} (\bibinfo {year}
  {2018})}\BibitemShut {NoStop}%
\bibitem [{\citenamefont {Scharf}\ \emph {et~al.}(2019)\citenamefont {Scharf},
  \citenamefont {Pientka}, \citenamefont {Ren}, \citenamefont {Yacoby},\ and\
  \citenamefont {Hankiewicz}}]{Scharf2019}%
  \BibitemOpen
  \bibfield  {author} {\bibinfo {author} {\bibfnamefont {B.}~\bibnamefont
  {Scharf}}, \bibinfo {author} {\bibfnamefont {F.}~\bibnamefont {Pientka}},
  \bibinfo {author} {\bibfnamefont {H.}~\bibnamefont {Ren}}, \bibinfo {author}
  {\bibfnamefont {A.}~\bibnamefont {Yacoby}},\ and\ \bibinfo {author}
  {\bibfnamefont {E.~M.}\ \bibnamefont {Hankiewicz}},\ }\bibfield  {title}
  {\bibinfo {title} {{Tuning topological superconductivity in phase-controlled
  Josephson junctions with Rashba and Dresselhaus spin-orbit coupling}},\
  }\href {https://doi.org/10.1103/PhysRevB.99.214503} {\bibfield  {journal}
  {\bibinfo  {journal} {Phys. Rev. B}\ }\textbf {\bibinfo {volume} {99}},\
  \bibinfo {pages} {214503} (\bibinfo {year} {2019})}\BibitemShut {NoStop}%
\bibitem [{\citenamefont {Fornieri}\ \emph {et~al.}(2019)\citenamefont
  {Fornieri}, \citenamefont {Whiticar}, \citenamefont {Setiawan}, \citenamefont
  {Portol{\'e}s}, \citenamefont {Drachmann}, \citenamefont {Keselman},
  \citenamefont {Gronin}, \citenamefont {Thomas}, \citenamefont {Wang},
  \citenamefont {Kallaher}, \citenamefont {Gardner}, \citenamefont {Berg},
  \citenamefont {Manfra}, \citenamefont {Stern}, \citenamefont {Marcus},\ and\
  \citenamefont {Nichele}}]{Fornieri2019}%
  \BibitemOpen
  \bibfield  {author} {\bibinfo {author} {\bibfnamefont {A.}~\bibnamefont
  {Fornieri}}, \bibinfo {author} {\bibfnamefont {A.~M.}\ \bibnamefont
  {Whiticar}}, \bibinfo {author} {\bibfnamefont {F.}~\bibnamefont {Setiawan}},
  \bibinfo {author} {\bibfnamefont {E.}~\bibnamefont {Portol{\'e}s}}, \bibinfo
  {author} {\bibfnamefont {A.~C.~C.}\ \bibnamefont {Drachmann}}, \bibinfo
  {author} {\bibfnamefont {A.}~\bibnamefont {Keselman}}, \bibinfo {author}
  {\bibfnamefont {S.}~\bibnamefont {Gronin}}, \bibinfo {author} {\bibfnamefont
  {C.}~\bibnamefont {Thomas}}, \bibinfo {author} {\bibfnamefont
  {T.}~\bibnamefont {Wang}}, \bibinfo {author} {\bibfnamefont {R.}~\bibnamefont
  {Kallaher}}, \bibinfo {author} {\bibfnamefont {G.~C.}\ \bibnamefont
  {Gardner}}, \bibinfo {author} {\bibfnamefont {E.}~\bibnamefont {Berg}},
  \bibinfo {author} {\bibfnamefont {M.~J.}\ \bibnamefont {Manfra}}, \bibinfo
  {author} {\bibfnamefont {A.}~\bibnamefont {Stern}}, \bibinfo {author}
  {\bibfnamefont {C.~M.}\ \bibnamefont {Marcus}},\ and\ \bibinfo {author}
  {\bibfnamefont {F.}~\bibnamefont {Nichele}},\ }\bibfield  {title} {\bibinfo
  {title} {Evidence of topological superconductivity in planar josephson
  junctions},\ }\href {https://doi.org/10.1038/s41586-019-1068-8} {\bibfield
  {journal} {\bibinfo  {journal} {Nature}\ }\textbf {\bibinfo {volume} {569}},\
  \bibinfo {pages} {89} (\bibinfo {year} {2019})}\BibitemShut {NoStop}%
\bibitem [{\citenamefont {Banerjee}\ \emph {et~al.}(2022)\citenamefont
  {Banerjee}, \citenamefont {Lesser}, \citenamefont {Rahman}, \citenamefont
  {Wang}, \citenamefont {Li}, \citenamefont {Kringh{\o}j}, \citenamefont
  {Whiticar}, \citenamefont {Drachmann}, \citenamefont {Thomas}, \citenamefont
  {Wang}, \citenamefont {Manfra}, \citenamefont {Berg}, \citenamefont {Oreg},
  \citenamefont {Stern},\ and\ \citenamefont {Marcus}}]{Banerjee2022}%
  \BibitemOpen
  \bibfield  {author} {\bibinfo {author} {\bibfnamefont {A.}~\bibnamefont
  {Banerjee}}, \bibinfo {author} {\bibfnamefont {O.}~\bibnamefont {Lesser}},
  \bibinfo {author} {\bibfnamefont {M.~A.}\ \bibnamefont {Rahman}}, \bibinfo
  {author} {\bibfnamefont {H.~R.}\ \bibnamefont {Wang}}, \bibinfo {author}
  {\bibfnamefont {M.~R.}\ \bibnamefont {Li}}, \bibinfo {author} {\bibfnamefont
  {A.}~\bibnamefont {Kringh{\o}j}}, \bibinfo {author} {\bibfnamefont {A.~M.}\
  \bibnamefont {Whiticar}}, \bibinfo {author} {\bibfnamefont {A.~C.~C.}\
  \bibnamefont {Drachmann}}, \bibinfo {author} {\bibfnamefont {C.}~\bibnamefont
  {Thomas}}, \bibinfo {author} {\bibfnamefont {T.}~\bibnamefont {Wang}},
  \bibinfo {author} {\bibfnamefont {M.~J.}\ \bibnamefont {Manfra}}, \bibinfo
  {author} {\bibfnamefont {E.}~\bibnamefont {Berg}}, \bibinfo {author}
  {\bibfnamefont {Y.}~\bibnamefont {Oreg}}, \bibinfo {author} {\bibfnamefont
  {A.}~\bibnamefont {Stern}},\ and\ \bibinfo {author} {\bibfnamefont {C.~M.}\
  \bibnamefont {Marcus}},\ }\href {http://arxiv.org/abs/2201.03453} {\bibinfo
  {title} {{Signatures of a topological phase transition in a planar Josephson
  junction}}} (\bibinfo {year} {2022}),\ \Eprint
  {https://arxiv.org/abs/2201.03453} {arXiv:2201.03453} \BibitemShut {NoStop}%
\end{thebibliography}

%

\clearpage
\newpage
\beginsupplement

\onecolumngrid
\begin{center}
\textbf{\Large Supplementary Information} 

\vspace{0.5cm}

\noindent \textbf{\large Sign reversal of the AC and DC supercurrent diode effect and  0--$\pi$-like transitions in ballistic Josephson~junctions}
\end{center}

\section{$L_0$ and  $L_0^{\prime}$ versus $B_y$ for both $B_y$ polarities.}

In Figs.~2\textbf{b,c} of the main text, we show the in-plane field ($B_y$) dependence of the first coefficients of the second-order-polynomial expansion of $L(I)$, namely, the Josephson inductance as a function of the current bias. For clarity, in that plot we have only shown the positive $B_y$ part of the graph. Sample 3 was indeed measured for both $B_y$-polarities and here we report and discuss the complete graph. 

\begin{figure*}[b]
\centering
\includegraphics[width=\textwidth]{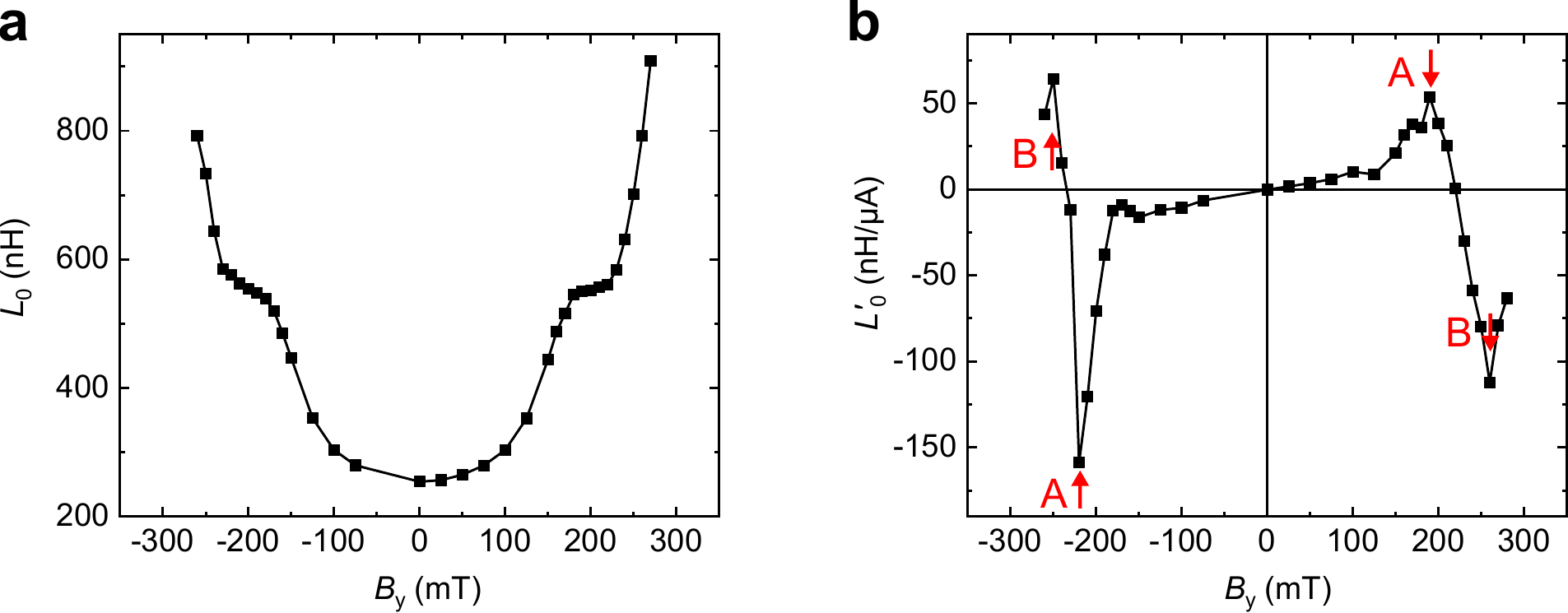}
\caption{\textbf{a},~ The coefficient $L_0$ (see text) and  \textbf{b}, the coefficient $L_0^{\prime}$,  plotted as a function of the in-plane magnetic field $B_y$ for both $B_y$-polarities.  The measurements were performed on sample 3. In the latter plot, the most relevant features are highlighted. These are the peaks in the magnetochiral anisotropy (A) and the sharp dips (B) where the sign of the magnetochiral anisotropy is inverted.}
\label{fig:firstSI}
\end{figure*}

Figure~\ref{fig:firstSI}\textbf{a} shows the $B_y$-dependence of $L_0$ for sample 3, which represents the zero-bias differential inductance. We notice that the graph appears approximately symmetric, although the negative plateau occurs at slightly higher $B_y$ and is more smeared. Figure~\ref{fig:firstSI}\textbf{b} shows the corresponding graph for $L_0^{\prime}$. In this case, the asymmetry between positive and negative $B_y$ is more pronounced. In particular, the peak A is much more evident for $B_y<0$. The  peak A [maximum of the AC supeconducting diode effect (SDE)]  occurs at $B_y=190$~mT for $B_y>0$ and at $B_y=-220$~mT for $B_y<0$. The peak B (maximum reversal of the AC SDE)  occurs at $B_y=260$~mT for $B_y>0$ and at $B_y=-250$~mT for $B_y<0$.

We cannot conclusively determine with certainty whether such differences must be attributed to experimental details, or whether they reflect physical asymmetries. Our simple theoretical model clearly does not break the symmetry between $B_y<0$ and $B_y>0$, since it assumes a perfectly isotropic Fermi surface, as it is the case for pure Rashba spin-orbit interaction (SOI). If real, these asymmetries might be the signature of Dresselhaus + Rashba SOI~\cite{BaumgartnerSI2022}, which for an arbitrary direction of the current might lead to small asymmetries in the AC SDE for the two polarities of the in-plane field.

\clearpage
\newpage

\section{Advantages of the array configuration}

At first glance, the most striking characteristics of our devices is the fact that they consist of a series of many (2250 for the devices reported here) junctions in series. However, in a previous work~\cite{baumgartner2020}, we have demonstrated that the physics probed by inductance measurements in such arrays is that of single junctions.

The reason why we work with long arrays and not with single junctions is merely technical. In fact, to make the interpretation of the experimental results as simple as possible, we operate in the low-frequency limit, i.e., the working point frequency of the RLC circuit $f_0=(2\pi\sqrt{L_eC_e})^{-1}$ is chosen to be much smaller than any physical resonance as, e.g., the Josephson-junction plasma frequency (240~GHz) or the transmission-line resonance (240~MHz for our array). Low frequencies make it also easier to operate under high magnetic fields, as required by our experiment. On the other hand, it would be extremely difficult to detect the typically small changes in the inductance (about 100~pH) of a single Josephson junction in the few-MHz~regime. The fabrication of an array of many junctions in series brings thus two main advantages: first, the inductance that we need to measure is multiplied by the number of junctions~(2250 in our case); second, we average over the imperfections of individual junctions, so that our results reflect as much as possible the intrinsic behavior under study. At high bias, the inductance of the weakest junction(s) clearly diverges, which leads to a premature emergence of finite resistance~\cite{baumgartner2020}. However, since we mainly focus on inductance measurements, those weak junctions are, in practice, irrelevant as long as the DC current bias does not become too high, as demonstrated in Ref.~\cite{baumgartner2020}.

\section{In-plane field dependence of the induced gap}
As demonstrated in Ref.~\cite{baumgartner2020}, Josephson inductance measurements make it possible to extract in a  reliable fashion both the average transmission coefficient $\bar{\tau}$ and the induced gap. 
To do so, we start from the Furusaki-Beenakker current-phase relation (CPR),  which is given by~\cite{BeenakkerPRL91,DellaRocca2007,RMPGolubov}
\begin{equation}
I(\varphi)=I_0f(\varphi) \equiv I_0\frac{\bar{\tau}\sin \varphi \tanh \left[\frac{\Delta^{\ast}}{2k_BT}\sqrt{1-\bar{\tau}\sin^2\left(\frac{\varphi}{2}\right)} \right]}{2\sqrt{1-\bar{\tau}\sin^2\left(\frac{\varphi}{2}\right)}},
\label{eq:cprshortball}
\end{equation}
where $\Delta^{\ast}(T)$ is the induced superconducting gap of the proximitized 2DEG and $\bar{\tau}$ is the average transmission coefficient. The parameter $I_0$ corresponds to the critical current only for $\bar{\tau}=1$ and $T=0$.

By sweeping the current, we can measure $L(I)$, thus we can plot the experimental $L_0/L(I)$ versus $2\pi L_0 I/\Phi_0$, where $L_0=L(0)$ is the inductance at zero bias. If we restrict ourselves to the low temperature limit, we can fit this curve with the corresponding prediction of Eq.~\ref{eq:cprshortball}, which for $T\rightarrow 0$ only depends on $\bar{\tau}$~\cite{baumgartner2020}. This is because the inductance is directly related to the CPR via \begin{equation}
L(\varphi)\ \equiv\ \frac{V}{\dot{I}}\ =\ \frac {\Phi_0}{2\pi I_0f'(\varphi)},
\label{eq:JI}
\end{equation}
where $\varphi=\varphi(I)$ is a function that only depends on the parameter $\bar{\tau}$. It can be obtained by inverting Eq.~\ref{eq:cprshortball}.
Once $\bar{\tau}$ (and thus $f(\varphi)$ for $T\rightarrow 0$) is found, we can 
immediately deduce $I_0$ from $L_0$ via
\begin{equation}
L_0\ \equiv\  \frac {\Phi_0}{2\pi I_0f'(0)},
\label{eq:zerobiasL}
\end{equation}
where, at low temperature, $f'(0)$ is known if $\bar{\tau}$ has been determined.

Then, we can use the relation between $I_0$ and  $\Delta^{\ast}$  to find the latter, if the number of channels $N$ is known~\footnote{Alternatively, one can use the temperature dependence of the zero-bias inductance to extract $\Delta^{\ast}$, if the number of channels $N$ is not known~\cite{baumgartner2020}.} (e.g., from the Sharvin resistance in the normal state),
\begin{equation}
I_0=\frac{e\Delta^{\ast}}{\hbar}N \implies \Delta^{\ast}=\frac{\hbar I_0}{eN}.
\label{eq:I0Dstar}
\end{equation}

\begin{figure}[tb]
\centering
\includegraphics[width=0.5\textwidth]{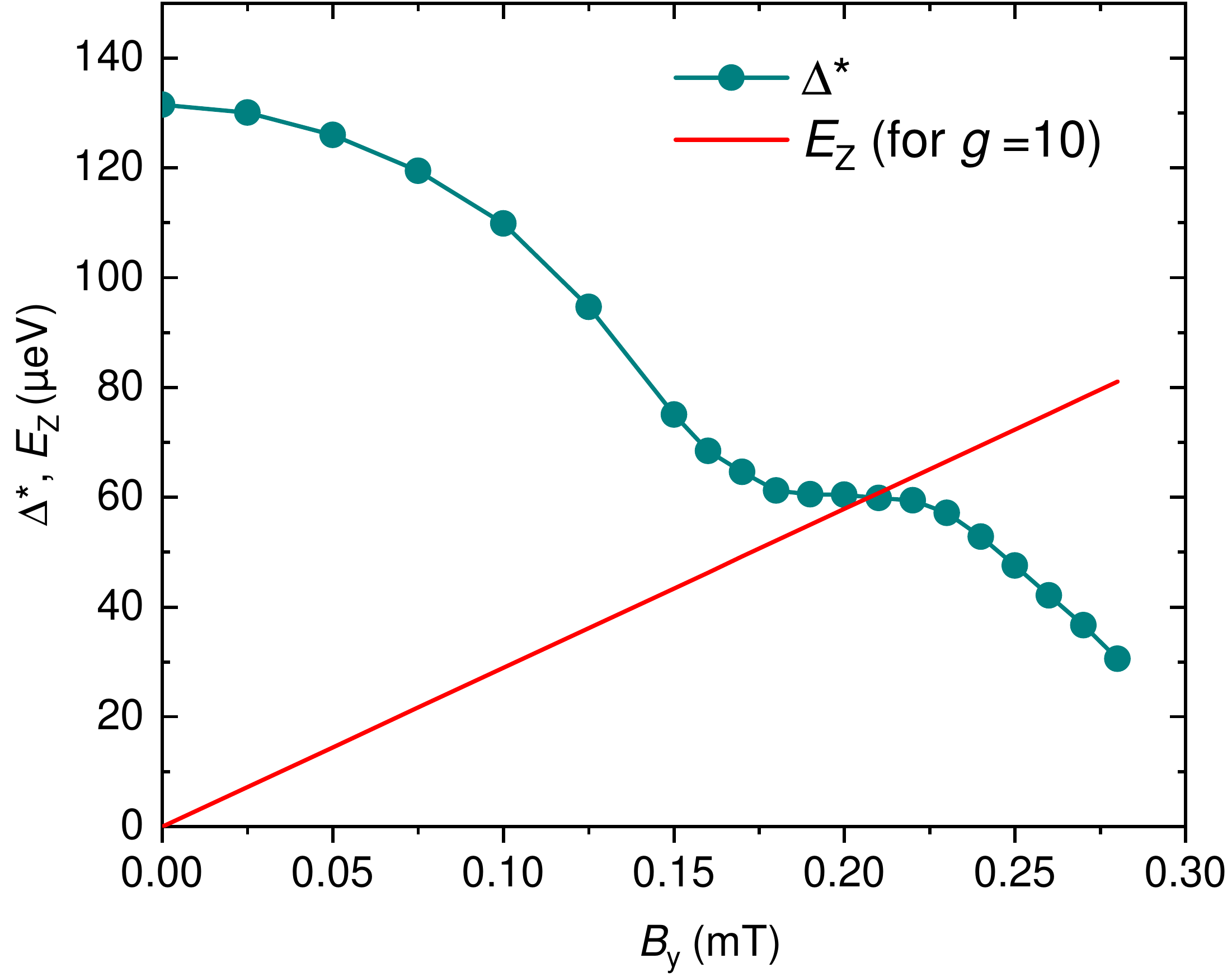}
\caption{In-plane magnetic field $B_y$-dependence of the superconducting gap induced in the 2D electron gas (dark cyan). The red line indicates the Zeeman energy for a $g$-factor of 10.}
\label{fig:deltastar}
\end{figure}

Combining the equations above, we can link directly $\Delta^{\ast}$ to $L_0$, so that we can extract $\Delta^{\ast}(B_y)$ from the measured $L_0(B_y)$, where the latter is directly measured. With that, we obtain the graph in Fig.~\ref{fig:deltastar}. We notice that the $B_y$-dependence of $\Delta^{\ast}$ is relatively flat for $B_y<50$~mT. 
At higher $B_y$-values, $\Delta^{\ast}$ significantly decreases. As discussed in the Methods section, the fact that our theory model ignores the $\Delta^{\ast}$-reduction makes it hard to quantitatively link the $\lambda_\mathrm{Z}$ parameter in the model with the in-plane field applied in the experiment, as we discussed in the Methods.

In Fig.~\ref{fig:deltastar}, we also plot the Zeeman energy $E_Z$ for $|g^{\ast}|=10$ (red line). Interestingly, the $0-\pi$-like transition [which is here witnessed by the plateau in $L(B_y)$, and thus in $\Delta^{\ast}(B_y)$] occurs when the two curves cross. 
In recent  theoretical and experimental works~\cite{Hart2017,Pientka2017,Costa2018,Scharf2019,Dartiailh2021} similar $0$-$\pi$ transitions were associated with a transition to a topological non-trivial phase, spectroscopically indicated by a zero-energy crossing of the lowest-energy Andreev bound states and a simultaneously modified ground-state parity---potentially resulting in topologically protected states in transversely confined systems~\cite{Scharf2019}. Even more recently, first spectroscopic evidence of such a topological transition was experimentally observed in the same system that we consider~\cite{Fornieri2019,Banerjee2022}. 
We note that both the details of our model and the parameter regime in which the $0$—$\pi$-like transitions occur in our theoretical study coincide with the topological regime discussed in Ref.~\cite{Scharf2019}.

\clearpage
\newpage
\section{Sign reversal of the DC SDE on another sample}
\begin{figure}[b]
\centering
\includegraphics[width=1\textwidth]{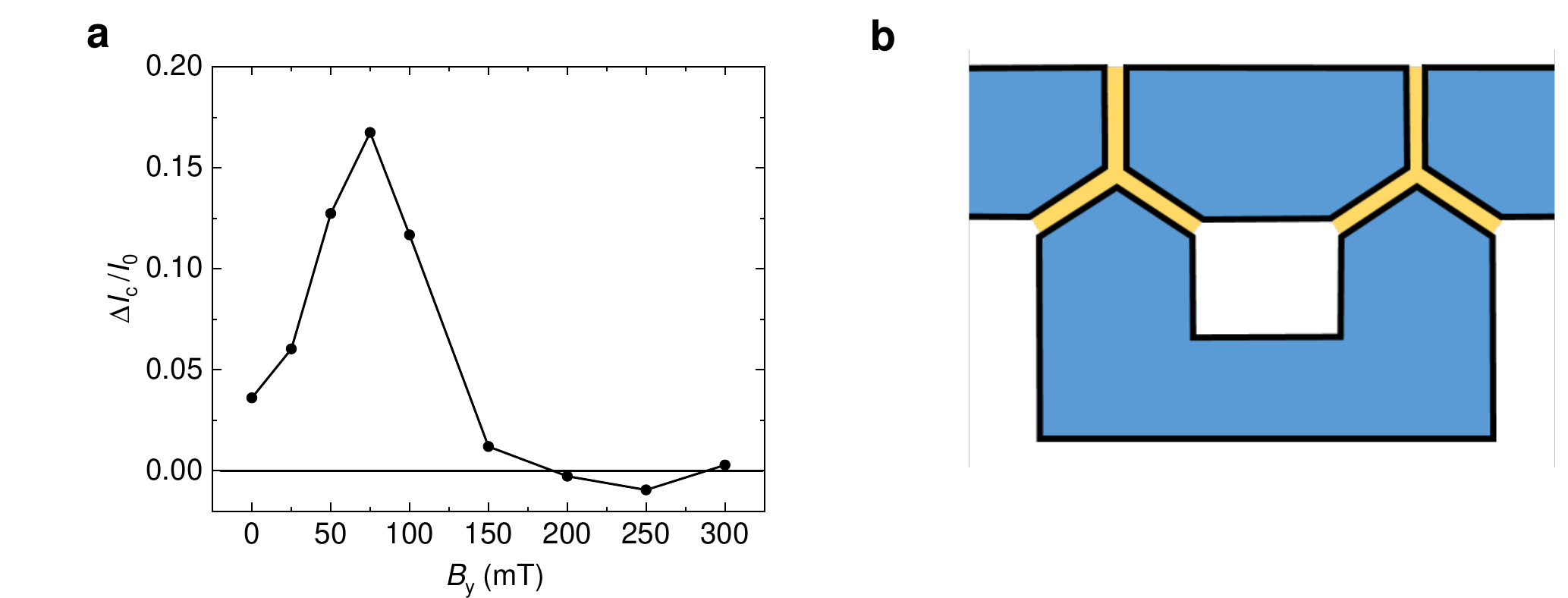}
\caption{\textbf{a}, In-plane field $B_y$-dependence of the critical current asymmetry $\Delta I_c$ measured in sample MTJJ4. The values are normalized to the average $I_0$ between positive and negative critical current at $B_y=0$, i.e., $I_0\equiv (I_c^+(0)+|I_c^-(0)|)/2$. \textbf{b}, Sketch of the unit cell of the array in sample MTJJ4. The light-blue regions correspond to a pristine Al/InAs heterostructure, the light yellow regions correspond to sample areas where the Al has been selectively etched, while the white regions correspond to sample parts where the quantum well is removed altogether by deep etching.}
\label{fig:trijun}
\end{figure}

Figure~\ref{fig:trijun}\textbf{a} shows the in-plane field $B_y$-dependence of $\Delta I_c$ for sample MTJJ4. While the wafer used to fabricate this sample is the same used  for samples 1 and 3, the device geometry is much different (see sketch in Fig.~\ref{fig:trijun}\textbf{b}) since this device is used for a different class of experiments, which are not within the scope of the present article. The device features an array of tri-junctions connected in pairs via loops. At zero out-of-plane fields, this device is basically a more complicated version of the Josephson junction arrays under study. In DC experiments, we can then measure $\Delta I_c(B_y)$: the results are similar to those reported for sample 1 in the main text, as one can deduce by comparing  the Fig.~\ref{fig:trijun}\textbf{a} here to  Fig.~3\textbf{b} of the main text. For the sample MTJJ4, the sign reversal of the DC SDE is more pronounced than in sample 1.

\clearpage

\end{document}